\documentclass[acmsmall,nonacm]{acmart}

\AtBeginDocument{%
  }

\setcopyright{cc}
\setcctype{by-nc-nd}
\acmJournal{PACMHCI}
\acmYear{2025} \acmVolume{9} \acmNumber{7} \acmArticle{405} \acmMonth{11} \acmPrice{}\acmDOI{10.1145/3757586}

\acmConference[Conference acronym 'XX]{Make sure to enter the correct
  conference title from your rights confirmation emai}{June 03--05,
  2018}{Woodstock, NY}
\acmISBN{978-1-4503-XXXX-X/18/06}

\begin{document}

\title[Dynamic Surveys]{Dynamic Surveys: Using LLMs to Blend Qualitative Depth, Quantitative Structure, and Collaborative Interaction}
\author{Kehua Lei}
\email{klei4@ucsc.edu}
\affiliation{%
  \institution{UC Santa Cruz}
  \country{USA}
}

\author{Aidan Ladenburg}
\email{aladenbu@ucsc.edu}
\affiliation{%
  \institution{UC Santa Cruz}
  \country{USA}
}

\author{Zahra Petiwala}
\email{zpetiwal@ucsc.edu}
\affiliation{%
  \institution{UC Santa Cruz}
  \country{USA}
}

\author{Zili Wang}
\email{zwang400@ucsc.edu}
\affiliation{%
  \institution{UC Santa Cruz}
  \country{USA}
}

\author{Dishita Jhawar}
\email{djhawar@ucsc.edu}
\affiliation{%
  \institution{UC Santa Cruz}
  \country{USA}
}

\author{Ipsita Bisht}
\email{ipsitabisht@gmail.com}
\affiliation{%
  \institution{UC Santa Cruz}
  \country{USA}
}

\author{Ansh Kumar}
\email{ansh.khalasi@gmail.com}
\affiliation{%
  \institution{UC Santa Cruz}
  \country{USA}
}

\author{David T. Lee}
\affiliation{%
  \institution{UC Santa Cruz}
  \country{USA}}
\email{dlee105@ucsc.edu}

\renewcommand{\shortauthors}{Kehua Lei et al.}

\begin{abstract}
  Surveys are a powerful tool for collecting data and eliciting insights on social phenomena, and are critical in product design, marketing, scientific research, and other domains. However, traditional open-ended and closed-ended question formats limit researchers’ ability to capture data that combines both the richness of qualitative insights and the analytical rigor of quantitative data. Closed-ended questions facilitate structured data collection that is amenable to statistical analysis but limit respondents' answers. In contrast, open-ended questions allow for nuanced responses incorporating new perspectives but require significant effort to interpret due to their unstructured nature. Moreover, traditional survey tools lack mechanisms to prompt respondents for deeper reflections or to facilitate engagement with others' perspectives, limiting the potential for richer insights. To address these problems, we propose Dynamic Surveys, a survey platform that uses Large Language Models (LLMs) to dynamically cluster qualitative responses in real time and to elicit quantitative ratings and rankings on those clusters and qualitative reflections on how their views compare to broader respondent trends, especially helpful in early-stage or exploratory research settings. This process generates a report showing survey creators and respondents the clustered responses as well as each cluster's rank, rating distribution, and follow-up reflections. To evaluate Dynamic Surveys, we conducted two field studies with 93 participants over a 2-month period. In the first study, 52 students provided input for a career workshop, while in the second, 41 students gave feedback on gaps in their academic curriculum. Of these, 44 respondents filled out a survey on their experience using Dynamic Surveys. We also shared the generated report with 4 individuals who were interested in the insights for their work, and interviewed them to understand their perspectives on the results and any     contextual risks they saw in the platform design. Our findings suggest that Dynamic Surveys not only provide richer and deeper insights into responses compared with traditional survey tools, but also increase engagement and foster a sense of community. We discuss broader implications for the design of survey platforms that blend qualitative depth with quantitative structure, facilitating richer insights and offering more collaborative interactions.
\end{abstract}



\received{October 2024}
\received[revised]{April 2025}
\received[accepted]{August 2025}

    \maketitle

\section{INTRODUCTION}

\begin{figure}[t]
 \centering
 \includegraphics[width=0.9\linewidth]{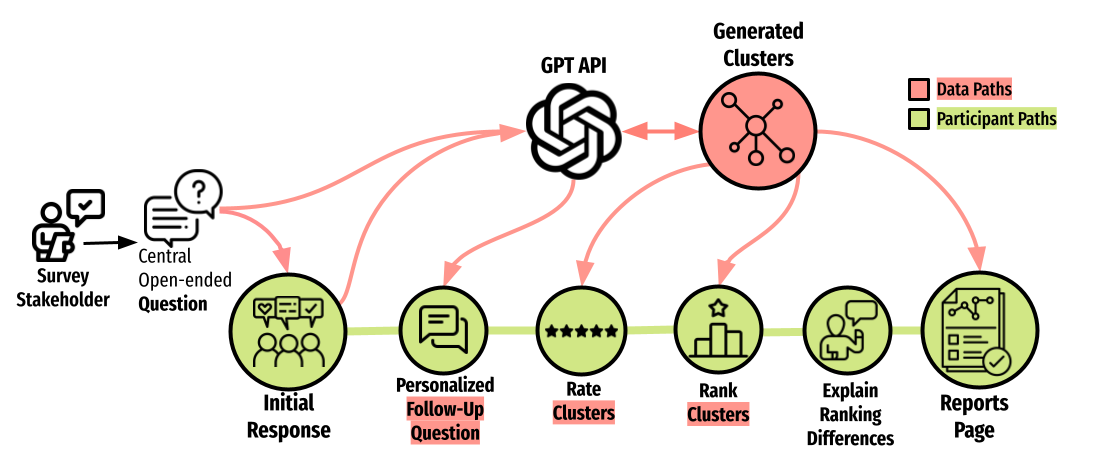}
 \caption{Workflow of Dynamic Surveys. The green flow at the bottom illustrates the respondent flow during the survey process. The upper section shows the back-end processes, including where ChatGPT APIs are called and how its clustering outputs are used. The two paths emerging from the GPT API represent the output of the two calls to the API for Personalized Follow-Up Question and Generated Clusters. Specific prompts for the clustering (Appendix A.1) and follow-up questions (Appendix A.2) are detailed in the appendix. }
 \label{fig:workflow}
\end{figure}

Surveys are a widely used tool for collecting data in diverse domains, from product design and marketing to social research and academic feedback. Their ability to gather insights quickly and cost-effectively, across broad geographic areas, makes them a popular choice for both researchers and practitioners~\cite{buchanan2009online, wright2005researching, parker1992collecting, mehta1995comparing}. However, traditional survey formats present significant challenges when it comes to capturing both the richness of qualitative data~\cite{rouder2021all, ferrario2022eliciting} and the analytical rigor of quantitative data~\cite{boone2005rigour}. Quantitative questions, such as multiple-choice or Likert scales, provide structured data that is easily analyzed but limit respondents' ability to fully express nuanced perspectives or to add new ones. Conversely, while open-ended qualitative questions encourage richer, more detailed insights, the resulting unstructured data is labor-intensive to process and hard to extrapolate to broader populations.

Just as importantly, traditional survey platforms operate in a one-way fashion, where respondents answer questions in isolation without the opportunity to interact with the perspectives of others. This lack of interaction limits the ability of researchers to foster community-driven insights and meaningful conversations among participants. Early innovations such as Wikisurveys~\cite{salganik2015wiki} have explored new ways of blending the strengths of diverse research methods while also enhancing respondent engagement, e.g. by allowing participants to rank on responses of others through pairwise comparisons. Related efforts have been made in the context of systems for participatory democracy~\cite{polis, kriplean2012supporting}. However, these methods tend to focus on voting at scale and thus stop short of fully integrating the richness of qualitative input with a dynamic, collaborative process. Wikisurveys demonstrated the potential to deepen engagement by making respondents active participants in the evolving insights, but little work has since expanded on this concept to build a more robust and interactive form of qualitative-quantitative data collection.

To address these limitations, we designed and implemented a survey platform, Dynamic Surveys, that builds on the community-driven potential of Wikisurveys while blending qualitative depth and quantitative structure in a more interactive and collaborative way. Dynamic Surveys is particularly useful for early-stage and exploratory research when deep interpretive immersion is not feasible and when the goal is not to build formal theory but to identify directional insights, surface common patterns, or provoke further inquiry, e.g. for systems-focused researchers and industry researchers looking to understand initial needs or experiences (see \textbf{Section \ref{sec:comparison}} for an in-depth comparison).

Dynamic Surveys uses large language models (LLMs) to elicit richer responses through generated follow-up questions, to dynamically cluster qualitative responses, and to ask respondents to reflect on, rate, rank, and engage with these clusters in real time. 
This interactive process transforms the survey from a solitary task into a community conversation, where participants actively shape the collective understanding, with the goal of fostering a greater sense of ownership over the results and of surfacing more nuanced community-driven insights that traditional survey tools might miss. Respondents can view the resulting survey report along with survey creators, which displays a ranked list of the clusters along with opinion distributions, summaries, detailed responses, and explanations of minority opinions. 

We evaluated the effectiveness of Dynamic Surveys through two field studies. The first involved 52 students providing input on discussion topics for a career workshop, and the second gathered feedback from 41 students on perceived gaps in their academic curriculum. Of these 93 participants, 44 filled out a survey on their experience using Dynamic Surveys. We also shared the generated report with 4 individuals who were interested in the generated insights for their work (who we will refer to as \textit{survey stakeholders}), and interviewed them to understand their perspectives on the results and any limitations or risks they saw in the viability of the platform design.

We found that Dynamic Surveys offered significant benefits to the data collection process. Survey stakeholders expressed that the platform helped them to efficiently understand collective perspectives and obtained valuable insights from the emergent clusters. Survey participants reported sharing more thoughtful responses compared to traditional survey platforms. Features like the follow-up questions, ranking comparisons, and the public reports page, helped participants feel more deeply involved while also fostering a sense of community. We conclude with a discussion on how future research can build on this work to introduce future survey tools that enhance collaborative interactions and blend strengths of diverse data collection methods.

Our paper makes the following three contributions: (1) the design of a novel survey platform that advances the community-driven potential of earlier work like Wikisurveys, while blending qualitative and quantitative data collection, demonstrating how a dynamic clustering capability powered by LLMs can create a novel “social layer” to elicit richer insights from survey participation; (2) insights from field studies that demonstrate how Dynamic Surveys fosters deeper respondent engagement and ownership of research outcomes; and (3) design implications for future survey tools that seek to enhance collaborative interactions and community-driven research processes, and that opens up methodological implications for lightweight, scalable elicitation and analysis of free-text responses in ways where theme generation and reflection happen collectively and asynchronously at scale.

\section{RELATED WORK}

\subsection{Survey Tools for Research}
Several commercial survey platforms, such as SurveyMonkey, Google Forms, and Qualtrics, are widely used for collecting data in qualitative research, especially particularly when deep interpretive immersion is not feasible and when the goal is primarily to elicit directional insights around feedback, experiences, and implications for iterative design in academic and applied settings. These tools offer customizable question designs and survey reporting features to varying extents; however, access to more advanced features often requires a paid subscription. Studies have shown that personalized survey designs can increase response rates~\cite{booker2021survey, munoz2010improving}, and respondents are more motivated to answer sensitive questions that relate to their personal experiences~\cite{griggs2018impact, thomas2024methodological}. Common survey tools, such as Qualtrics, offer embedded data features to support logic branching and more personalized question designs based on recorded data. Additionally, researchers have explored data-driven survey designs that integrate diverse data sources, allowing for more personalized question tailoring and broadening survey applications across different fields. For example, surveys have been developed using health data from wearable devices~\cite{heritier2023food}, location data from applications~\cite{huguenin2017predictive}, and local language, Pidgin English and animated GIFs~\cite{hannah2021interactive}. Recently, new platforms has been developed to facilitate the import of third-party data~\cite{velykoivanenko2024designing} and support external web content embedding~\cite{ebert2023qbutterfly}.

Additionally, a body of research has developed tools and online platforms to structure the collection and analysis of qualitative data across a range of fields. In psychology, for instance, the Visual Analogue Scale (VAS)~\cite{reips2008interval} and the Likert Scale~\cite{joshi2015likert} are widely used in surveys to capture subjective experiences. In economics, text mining techniques have been applied to policy documents and public speeches to extract and quantify qualitative information~\cite{gentzkow2019text}, while prediction markets employed market mechanisms to aggregate individual qualitative assessments into collective quantitative forecasts~\cite{wolfers2004prediction}. 

Studies also suggest that interactive surveys with real-time feedback can lead to higher-quality responses, and recent research has begun exploring how surveys can support greater interactivity. Recent research has begun exploring how surveys can support greater interactivity. For instance, SMARTRIQS groups respondents based on predefined conditions, enabling them to participate in voting or real-time discussions~\cite{molnar2019smartriqs}. Other studies have examined the use of chatbots in surveys~\cite{zarouali2024comparing, rhim2022application}. For example, Wen and Colley proposed a hybrid survey system in which a chatbot intervenes with respondents who intend to skip questions~\cite{wen2022hybrid}. Xiao et al. found that chatbot-driven surveys can elicit more relevant and specific responses compared to conventional survey tools~\cite{xiao2020tell}.

Interactive modes for data collection have also been explored in research on online deliberation. While these projects tend to focus on structuring public opinion to support decision-making and to facilitate public discourse, they also provide models for interactively eliciting and synthesizing rich data from participants at scale. For example, Consider.it facilitated deliberation by allowing participants to annotate pros and cons for each topic, converting these inputs into metrics of support and opposition percentages and visualizations of participant distributions across various positions~\cite{kriplean2012supporting}. Voting mechanisms are also  widely adopted to quantify public opinion. LiquidFeedback, for instance, employed a dynamic delegation system to transform qualitative opinions into collective quantitative decisions~\cite{behrens2014principles}, while Pol.is enabled participants to express agreement, disagreement, or neutrality toward statements, using cluster analysis to reveal areas of consensus and divergence~\cite{polis}. Similarly, Wikisurveys employed pairwise voting to crowdsource ideas for public policy proposals and organizes feedback in a prioritized ranked list based on collective preferences~\cite{salganik2015wiki}. Though this study focuses on supporting elicitation of rich insights rather than consensus-building and deliberation, we draw from these interactive mechanisms to engage respondents in reacting to or building on others' responses in ways that enrich the derived insights. Prior work has shown that participants can contribute meaningfully to the interpretation of open-ended data through extended processes involving multiple rounds of reflection and discussion~\cite{baumer2017subjects}. In contrast, our work explores how such interpretive and participatory dynamics might be embedded more directly within the survey process itself. Instead of requiring separate rounds of coding or synthesis, Dynamic Surveys integrate collaborative input such as clustering, reflection, and ranking into the flow of data collection. This approach offers a more lightweight and scalable alternative to multi-stage designs. In the Discussion \textbf{(Section \ref{LLM discussion})}, we discuss opportunities to build on this work in future studies to explore their potential for consensus-building in settings where deliberation is not just a nice-to-have for eliciting richer insights, but critical for harmonizing conflicting viewpoints and perspectives towards an understanding of the issues.

\subsection{AI-augmented Qualitative Data Analysis}
Coding and analyzing unstructured text data from participants is a vital step in qualitative research, providing in-depth, context-driven insights~\cite{denzin1996handbook} that allow researchers to view issues from participants’ perspectives~\cite{maxwell2013qualitative}. However, this process is often time-consuming and error-prone~\cite{cypress2019data}. Qualitative data analysis (QDA) tools such as NVivo, MAXQDA, and ATLAS.ti are widely used by researchers to support coding and analysis~\cite{freitas2018case}, streamlining the coding process and facilitating the organization of coded data.

Beyond existing commercial QDA tools, a growing body of research has explored the use of AI to better support large-scale coding tasks. Researchers have applied topic modeling to qualitative analysis, enabling early exploratory analysis by identifying broad themes in large text-based datasets~\cite{wesslen2018computer, andreotta2019analyzing, bakharia2016interactive, lennon2021developing}. Since coding remains a foundational yet resource-intensive step in qualitative analysis~\cite{castleberry2018thematic}, efforts have been made to simplify annotation by leveraging machine learning to support coding through pre-trained models~\cite{stenetorp2012brat, crowston2010machine, li2021qualitative, guetterman2018augmenting}. To improve the recall and accuracy of these models, semi-automated coding systems have been developed that iteratively learn from user inputs during the coding process~\cite{lejeune2011normal, yan2014semi, marathe2018semi}. However, the results of these AI-assisted analyses often lack interpretability and exploratory value~\cite{bazeley2009analysing, grimmer2013text} and are limited in providing the depth of insights that human researchers bring. Moreover, AI-based coding models typically lack the flexibility that human coders employ to iteratively refine codes, resulting in limited outcomes~\cite{chen2018using, feuston2021putting}. Thus, researchers have started to develop more interactive coding systems, emphasizing that these AI-assisted systems should be interpretable to humans and should allow researchers to dynamically engage with and refine the model~\cite{chen2016challenges, klie2018inception, jiang2021supporting}. Such systems enable users to provide feedback on coding results, offer explanations for suggested codes to enhance transparency and interpretability~\cite{rietz2021cody, gebreegziabher2023patat}, or integrate visual interactions to support researchers in exploring connections between codes~\cite{hong2022scholastic, goldman2022quad}, thereby facilitating better human-machine collaboration.

Recently, the strong performance of large language models (LLMs) in understanding and generating human language has led to their increasing application in qualitative analysis ~\cite{dai2023llm}. Katz et al. used ChatGPT to analyze student team feedback by mapping it to a predefined taxonomy, while Zhang et al. explored prompt design for using ChatGPT to simplify thematic analysis tasks~\cite{zhang2023redefining}. CollabCoder~\cite{gao2024collabcoder} and CoAIcoder~\cite{gao2023coaicoder} further demonstrated how LLMs can not only assist in analysis but also enhance collaboration among researchers. Despite their strengths, researchers have noted that LLM-based analyses require validation and cannot fully replace human analysts~\cite{zhang2024qualitative}, as relying solely on LLMs may reduce analytical diversity~\cite{gao2023coaicoder}. Consequently, the most valuable use of LLMs lies in complementing human researchers’ interpretation~\cite{khan2024automating}.

Whereas most of this work focuses on applying AI to assist human analysts during the coding and analysis phase following data collection, our work explores how LLMs can support earlier stages of the research process, particularly the elicitation and organization of qualitative input during data collection. In this paper, we leverage LLMs’ strong performance in inductive reasoning to generate dynamic clustering and reflection prompts based on participant responses. Rather than simply coding after the fact, our system structures input as it is being gathered, enabling a lightweight ``social layer'' to surveys that invites participants to engage with and build on emerging themes. This approach has the potential to yield more informative and structured data for researchers while also facilitating collective sensemaking among respondents.

\section{DYNAMIC SURVEYS}

To explore survey designs that combine the richness of qualitative data collection with the structured nature of quantitative data in an interactive collaborative process, we developed Dynamic Surveys, an online survey platform built using an Angular, NgRx, Firebase, and Cloud Functions stack that integrates both qualitative and quantitative data through the use of large language models (LLMs). LLMs are used through calls to OpenAI APIs run on Google Cloud Functions. These generate follow-up questions that elicit richer insights and organize responses into thematic clusters for rating, ranking, and reflection in subsequent questions. In what follows, we walk through the user experience of a survey respondent as well as the technical details in the system implementation for supporting each step in the flow (\textbf{Figure \ref{fig:workflow}}). We end with a comparison of Dynamic Surveys against other methods for data collection, discussing when Dynamic Surveys might be uniquely valuable, e.g. for use by particular types of researchers or in particular research contexts.

\subsection{The Initial Survey Question and Response}

Dynamic Surveys centers on eliciting insights around a single focused research question. The process starts with a single focused open-ended question. For example, the questions in our studies were ``\textit{What question would you love to ask a university recruiter in tech?}'' and ``\textit{What do you feel are the gaps in the engineering courses/curriculum?}'' This question is provided by the survey creator in an interface similar to that of most survey platforms. When respondents go to fill out the survey, they see this single question on the first page and write out their response before clicking "Next" to move onto the next page.

\subsection{Eliciting Richer Stories Through a Follow-up Question}

On the next page, the platform calls the OpenAI GPT-4o API with a custom prompt along with their response to generate a tailored follow-up question. We designed this prompt to encourage deeper reflection and elaboration (see \textbf{Appendix A.1}), and refined it through a protostudy by testing it on sample open-ended responses. This ensured the generated questions were relevant, built appropriately on participants’ input, and prompted deeper reasoning, examples, or context. These follow-up questions aim to elicit richer input, similar to how an interviewer might encourage a respondent to elaborate in a conversation. For example, in the survey on questions to improve engineering curriculum, a student responded: "\textit{We have some good courses that match the skills of today's CS industry, but some courses are outdated. we need more courses matching the industry.}" They were then given the personalized follow-up question: \textit{"Which specific courses do you believe are outdated, and what new topics or skills would you recommend being added to better align with current industry demands?"}, to which they responded: \textit{"AI and NLP courses are outdated. We should add topics infusing ChatGPT or current established AI model[s], or topics creating our own chatbot"}. While the original response provided a general critique of outdated courses, the targeted follow-up question encouraged the respondent to articulate specific actionable recommendations for the survey stakeholder to implement. 

\subsection{Concurrent Clustering of Responses Into Themes}

While participants are writing a response to the follow-up question, the platform is simultaneously running a cloud function to extract the themes from the initial response into clusters. There are two main steps involved. First, the platform uses the OpenAI model GPT-4o to compare the response with existing clusters and add it into any cluster it aligns with. For the first response, before any clusters are formed, this step is skipped. Next, the model finds any core concepts in the response not covered by the existing clusters, and generates one or more new clusters to fill the gap. It is not uncommon for a response to contain more than one theme, so it's important for our system to be able to categorize responses into more than one cluster. We designed a structured prompt that guides the model to generate clusters with clear, concise, and thematically meaningful names that are distinct from one another, while avoiding redundancy or unnecessary fragmentation (see \textbf{Appendix A.2}). For example, a sample response from the career workshop survey: "\textit{some things that come to mind are what the work environment/culture is like and how do companies typically go through the interview/training process}" could be simultaneously put into the existing cluster: "\textit{Work Environment In Tech Companies}" while also generating a new cluster: "\textit{Interview And Training Processes}".

\begin{figure}[b]
  \centering
  \includegraphics[width=\linewidth]{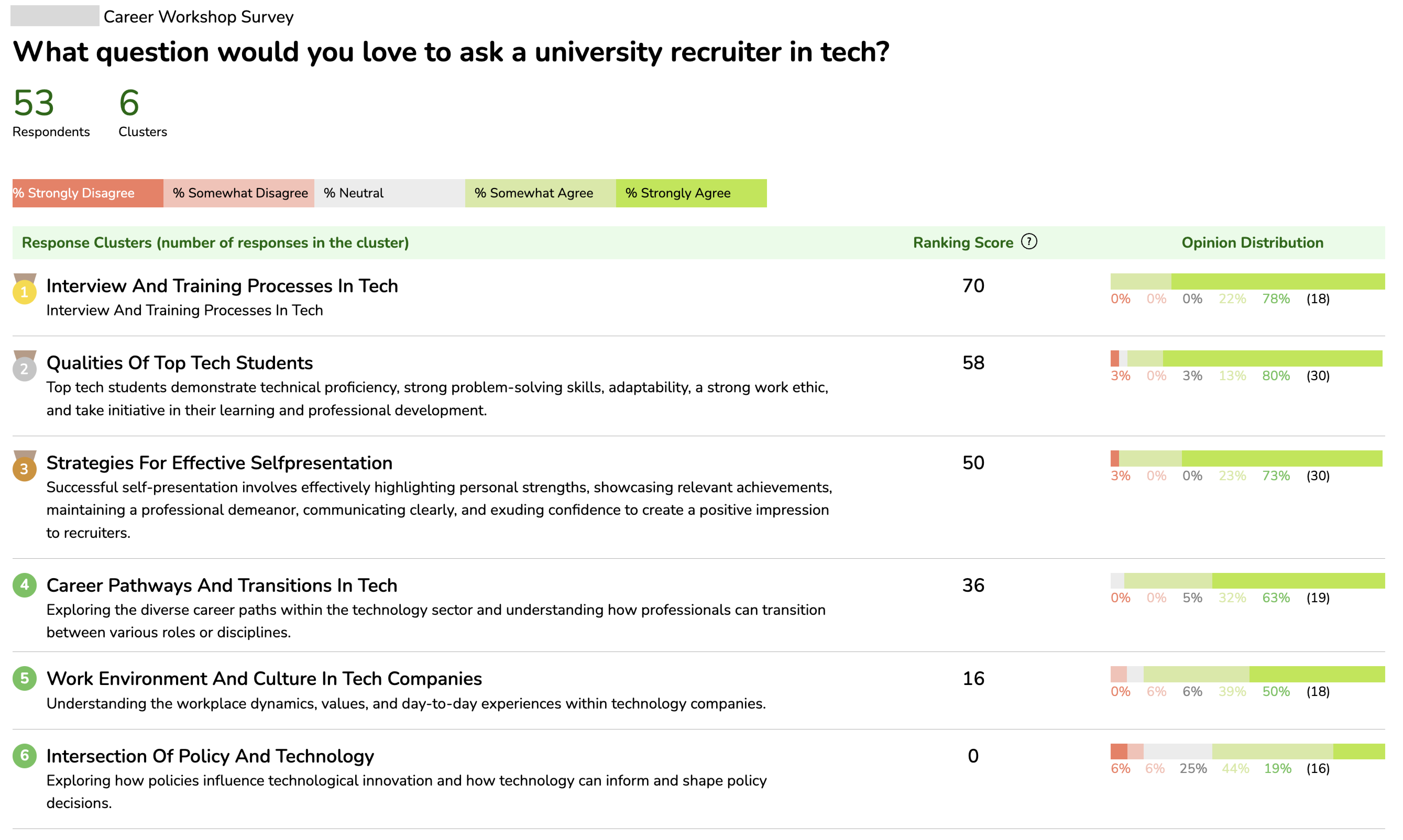}
  \caption{Screenshot of the first screen of the Career Workshop Survey, displaying cluster rankings, each cluster's ranking score, and opinion distribution. Certain details have been obscured to maintain the anonymity required for this paper.}
  \label{fig:report_ranking}
\end{figure}

\subsection{Rating and Ranking of Theme Clusters}

By the time respondents write their response to the follow-up question and click "Next", the updated theme clusters will have been generated. Participants are then brought to a page in which they are asked to rate each cluster on a 5-point Likert scale, from “Strongly Agree” to “Strongly Disagree” with an additional “N/A” option. These aim to assess the extent to which the identified clusters are shared across the respondent pool, e.g. to what degree others in the community experienced the same issues raised, have the same preferences, and so on. For example, in the survey on career workshop topics, this page asked respondents: "\textit{Reflecting on your own goals and experiences, how strongly do you agree or disagree that the following topics would be helpful for you to hear from the recruiter?}".

Upon submitting their Likert-scale ratings, they are brought to another page in which they are asked to rank the clusters based on its value for the group as a whole. This is slightly different from the Likert-scale question which focused on personal experiences. For example, one might rank something high because they believe its an important topic for a career workshop even if it's not something they have personally had questions about. Eliciting ordinal ranking data also has the added benefit of forcing respondents to prioritize (e.g. in cases when everything is rated with similar levels of importance).

\begin{figure}[t]
  \centering
  \includegraphics[width=\linewidth]{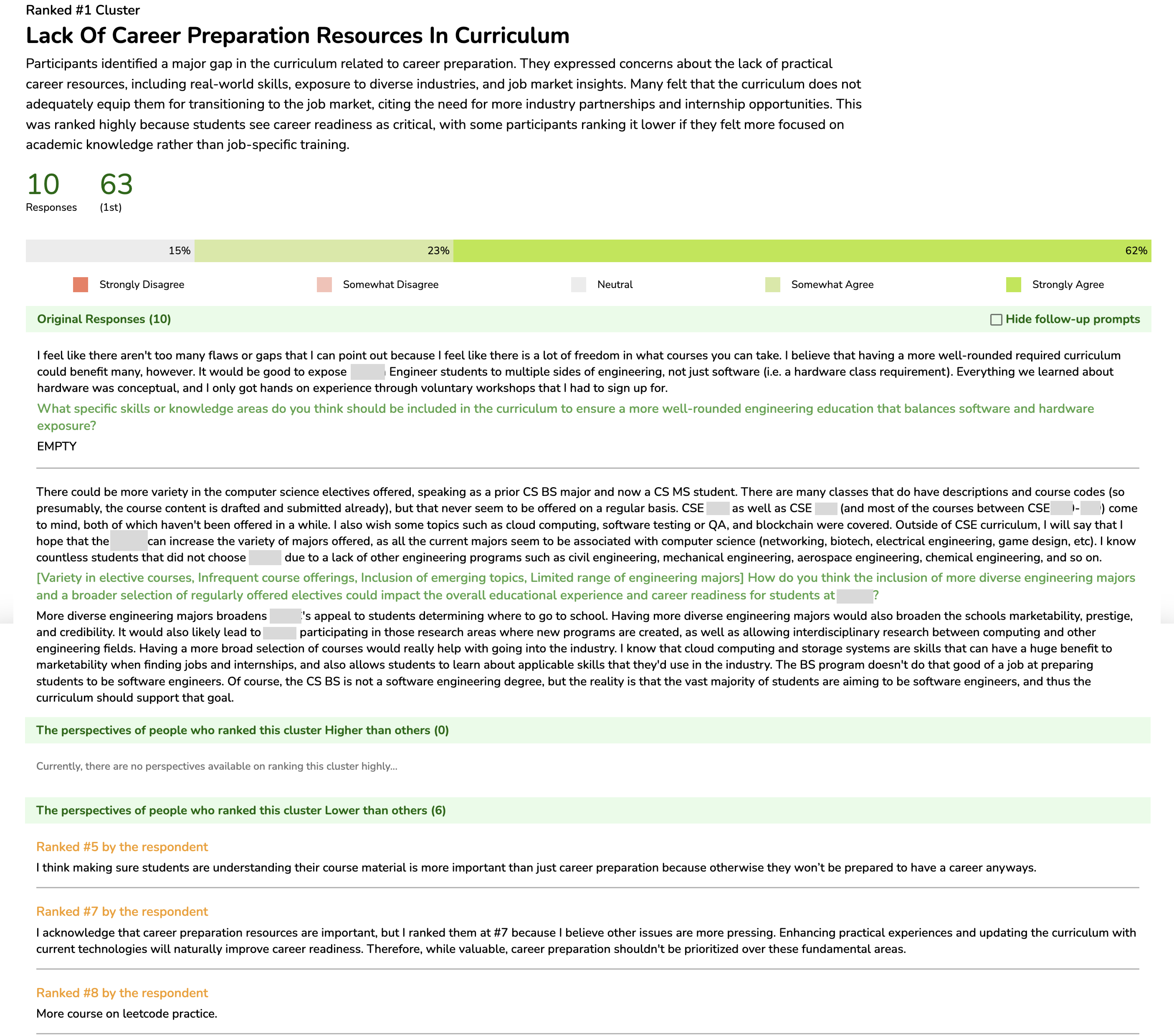}
  \caption{Screenshot from the Department Course Curriculum Survey showing details of the first cluster. It includes a GPT-generated summary highlighting key insights shared by respondents, along with all original responses, follow-up questions with answers, and explanations from respondents on why they ranked the cluster higher or lower than others. Certain details have been obscured to maintain the anonymity required for this paper.}
  \label{fig:report_cluster}
\end{figure}

\subsection{Explaining Differences Between Personal and Aggregate Preferences}

Finally, after submitting ranking preferences, respondents are shown a comparison of their ranking with the aggregated ranking from other respondents and are asked to explain any significant differences, i.e. \textit{"Below, we've listed areas for the topics that you ranked really high or low, but in ways that differed from others. Can you share your experiences or contexts to help others understand your reasons behind your rankings?"} They are then shown clusters that they ranked high but others ranked low, and vice versa, and given a chance to explain further. These questions provide an opportunity for further reflection and for sharing unique lived experiences that might help others to better understand minority perspectives.

\subsection{The Reports Page: A Synthesized Summary and Cluster Details}

These insights are aggregated together in a survey report page which can be viewed by both survey creators and respondents. The report page starts with a summary view that displays clusters as a ranked list, with their title, description, ranking score, and opinion distribution. This enables viewers to easily see prioritized areas as well as points of agreement and divergence within the data (\textbf{Figure \ref{fig:report_ranking}}).

To compute the ranking score, we essentially used the Borda rule~\cite{chamberlin1983representative}, one of the most prominent voting rules, but with a slight modification of the score to de-emphasize themes that had not yet been seen and ranked by many people yet. The modification can intuitively be understood as a lower confidence bound of the score. If the theme has not been seen by many people, then there is a wide range in the confidence interval, resulting in a low lower confidence bound. Highly ranked clusters in the list should be understood as clusters for which we have high confidence in their high ranking score. Low ranked clusters, on the other hand, could be truly low ranking, or just not yet seen by enough people to have confidence in its ranking score.

The report also contains more detailed sections for each cluster underneath the summary view. These sections include the individual responses, the follow-up questions, responses to follow-up questions, and explanations from respondents whose rankings differed from the group (\textbf{Figure \ref{fig:report_cluster}}). We used GPT to generate a summary for each cluster.

\subsection{Comparing Dynamic Surveys with Other Research Methods}\label{sec:comparison}
Dynamic Surveys introduces a new approach to data collection that retains the scalability and lightweight accessibility of conventional survey tools, while encouraging respondents to reflect, elaborate, and engage with others’ perspectives. Traditional survey formats often present a tension between structure and expressiveness. Closed-ended questions support statistical analysis but limit elaboration, while open-ended responses offer nuance but result in unstructured data that is difficult to interpret. Moreover, both formats typically collect responses in isolation, without mechanisms for participants to build on others’ input. Dynamic Surveys illustrate how interactive survey-based systems can support reflection and comparison in ways that offer some of the interpretive affordances of interviews and focus groups, while requiring less time and analytical expertise (see \textbf{Table \ref{tab:comparison}}).

\begin{table}
\caption{Comparison of Dynamic Surveys and Other Methods} 
\centering
\begin{tabular}{p{0.15\textwidth}|p{0.1\textwidth}|p{0.12\textwidth}|p{0.1\textwidth}|p{0.1\textwidth}|p{0.1\textwidth}|p{0.1\textwidth}}
\toprule

Method & Time \newline Efficiency & Insight Depth & Scalability & Required Expertise & Personali-zation & Interaction \\
\hline
Closed-ended Survey & High & Low & High & Low & None & None \\
\hline
Open-ended Survey & High & Moderate & High & Moderate & None & None \\
\hline
Interview & Low & High & Low & High & High & High \\
\hline
Focus Group & Low & High & Low & High & High & High \\
\hline
Dynamic \newline Surveys & High & Moderately high & High & Low & Moderate & Moderate \\

\bottomrule

\end{tabular}
\label{tab:comparison}

\end{table}

Dynamic Surveys are particularly well suited for contexts where researchers or designers seek to elicit feedback, experiences, or reflections that extend beyond what traditional surveys typically afford and to understand how such themes resonate across a broader group. Rather than replacing interviews or quantitative surveys, Dynamic Surveys offer a lightweight system for gathering open-ended insight and assessing alignment or divergence among participants. They are especially useful when time, resources, or expertise in qualitative interpretation or statistical analysis are limited.

This makes them especially valuable for industry researchers and systems-focused practitioners conducting small studies similar to focus groups. For example, they can be used to gather early intuition on user needs during pilot stages or to collect post-use feedback on prototypes. They also support educators, community organizers, and facilitators who seek to engage broader participation without conducting full-scale qualitative analysis. Because surveys can be launched with a single open-ended prompt and require no prior training in qualitative methods, the system is accessible to users without formal research design experience. In participatory settings, the ability to cluster and compare responses allows broader community voices to surface. Dynamic Surveys can also assist expert researchers during early exploratory phases by helping to surface diverse perspectives, guide subsequent research directions, and inform the design of more targeted qualitative research.

While Dynamic Surveys support reflection and comparison across participant input, they are not designed to support the goals of interpretive qualitative paradigms such as grounded theory~\cite{charmaz2015grounded}, reflexive thematic analysis~\cite{braun2019reflecting}, participatory action research~\cite{baum2006participatory}, or critical and hermeneutic traditions~\cite{kinsella2006hermeneutics}. These paradigms emphasize sustained researcher immersion, iterative sense-making, and contextually grounded interpretation, which are often informed by researcher reflexivity, collaboration with participants, and critical engagement with broader social structures. Dynamic Surveys provide a practical tool for identifying patterns, surfacing perspectives, and provoking further inquiry in contexts where deep interpretive work may not be feasible or necessary. By organizing collective input and highlighting emergent themes, Dynamic Surveys also offer a novel system-based approach to early-stage insight generation, which may support exploratory theorizing in both academic and applied settings. While Dynamic Surveys provide some level of quantitative signal to help surface themes across respondents, they are not intended for studies requiring statistical generalization or experimental control. 

\vspace{\baselineskip}
\section{METHODS}

To understand participants' experiences with completing the Dynamic Surveys and the value of the data presented in the survey report, we conducted two field studies. A total of 97 participants were involved, including 52 who completed one survey and 41 participants who completed a different survey, with the 4 survey stakeholders (see \textbf{Table \ref{tab:interview participants}}) also participating in the study. 

This study was approved by the Institutional Review Board of the first author’s university. The IRB reviewed the study under standard exemption criteria and determined it posed minimal risk to participants. Participation was voluntary and participants could withdraw from the study at any time.

\begin{table}
\centering
\caption{Interview participant demographics and background. \protect\footnotemark } 
\begin{tabular}{p{0.06\textwidth}|p{0.10\textwidth}|p{0.2\textwidth}|p{0.27\textwidth}|p{0.13\textwidth}}
\toprule
ID & Gender & Job Title & Survey Name & Expertise in Qualitative Research \\
\hline
I1 & Female & Recruiter & Career Workshop Survey & Intermediate \\
\hline
I2 & Female & Department \newline Administrator & Department Course \newline Curriculum Survey & Expert \\
\hline
I3 & Female & Department \newline Administrator & Department Course \newline Curriculum Survey & Expert \\
\hline
I4 & Female & Department \newline Administrator & Department Course \newline Curriculum Survey & Intermediate \\
\bottomrule

\end{tabular}
\label{tab:interview participants}
\end{table}
\footnotetext{We chose not to specify the exact roles of the three department administrators to maintain anonymity. However, all three are actively involved in improving the academic experience within their department.}

\subsection{Study Procedures and Participants}
We evaluated Dynamic Surveys in two settings, one focused on career workshop preparation and the other on curriculum feedback. Both are examples of contexts where researchers or organizers may want to elicit rich input grounded in lived experiences, without requiring intensive facilitation or methodological expertise. These settings illustrate how Dynamic Surveys can support insight-oriented data collection in ways that surface shared themes while preserving individual perspectives.

Our goal in these studies was not to evaluate consensus-building or decision-making processes. Instead, we focused on how interactive survey dynamics can help surface insights that are meaningful to the communities involved. In the Discussion \textbf{(Section \ref{LLM discussion})}, we outline future directions for evaluating the potential of Dynamic Surveys in more deliberative settings, where navigating differing perspectives and fostering mutual understanding may play a more central role.
\subsubsection{Study 1: Career Workshop Panel Questions}
In the first study, we collected panel questions for an upcoming workshop focused on providing career guidance to students pursuing tech industry jobs after graduation. The workshop was organized by a recruiter from a tech company in collaboration with a university lab. After discussions with the workshop organizers, we created a Dynamic Survey with the question, "What question would you love to ask a university recruiter in tech?" We distributed the survey through the lab's Slack channel and a related course's Discord server. A total of 52 students participated in the Dynamic Survey, and 18 of them later completed a post-study survey via Google forms to provide feedback on their experience. These 18 participants were entered into a lottery for a \$20 Amazon gift card. After collecting responses to the Dynamic Survey, we first surveyed the recruiter and then conducted a 30-minute semi-structured interview to understand her views of the survey report.

\subsubsection{Study 2: Engineering Curriculum Gaps}
In the second study, we aimed to gather student feedback on improvement areas within the courses and curriculum offered by the engineering department at a university. We created a Dynamic Survey with the question, "What do you feel are the gaps in the engineering courses/curriculum?" We shared recruitment information via social media, and 41 students participated in the survey. Of these participants, 26 completed a post-study survey and were entered into a lottery for a \$20 Amazon gift card. We then surveyed and conducted 30-minute semi-structured interviews with 3 faculties from the department, who provided insights on the survey findings.

\subsection{Surveys}
Our recruitment advertisement included both the Dynamic Survey and the post-study survey.  Participants who completed the Dynamic Survey could voluntarily choose whether to fill out the post-study survey, incentivized by entry into a lottery for a gift card upon completion. The post-study survey asked participants about their likes and dislikes regarding the use of Dynamic Surveys for sharing ideas, and how their experience compared to other survey tools in terms of the overall survey process and the value of the report. All ratings of likelihood were done on a 5-point Likert scale. Participants were asked to rate how strongly they agreed with appreciating follow-up questions, rating and ranking others' responses, sharing deeper thoughts than in other feedback surveys, resonating with others' responses, gaining new perspectives, finding the clusters clear and coherent, and their experience with the length of the survey. Finally, participants were asked how likely they were to recommend Dynamic Surveys to others for collecting data.

Alongside the participants' post-study survey, we also surveyed the survey stakeholders before conducting semi-structured interviews with them. This post-study survey focused on their views of the survey report. We asked what they liked and disliked about using Dynamic Surveys to collect feedback, how Dynamic Surveys helped them understand participants’ perspectives, and what additional value it offered compared to other survey tools. They were also asked to rate, the clarity and coherence of the clusters, the value gained from follow-up responses and rankings, the added depth from rating and ranking clusters, and the ease of interpreting the generated report. At the end of the survey, we asked how likely they were to recommend Dynamic Surveys to others for collecting data.

\subsection{Semi-Structured Interviews}
Our semi-structured interviews lasted 30 minutes and were conducted virtually via Zoom video conferencing. We began by asking the survey stakeholders about their prior experiences with collecting and analyzing data using other survey tools, and how these compared to their experience with Dynamic Surveys. We then asked them to elaborate on their responses to the Likert scale statements from the post-study survey. Lastly, we explored the unique value Dynamic Surveys offered in helping them understand participants' insights compared to other tools.

\subsection{Data Analysis}
To evaluate the quality and value of the dynamic survey reports, we analyzed the accuracy of the clusters and examined the relationship between cluster rankings and the percentage of participants who selected "strongly agree" for the Likert question. Additionally, we used the GPT clustering prompts applied in the Dynamic Surveys platform to explore whether follow-up responses and ranking explanations could generate new themes from the original responses.

The interviews were recorded, transcribed, and analyzed along with the survey responses using an inductive thematic analysis approach. The first author began by independently open-coding one interview transcript and the first half of the post-study survey responses from the career workshop study, developing a codebook in the process. The first author then discussed the initial codes and codebook with another team member, making minor revisions to ensure a shared understanding before proceeding. The two researchers then used this revised codebook to collaboratively code the remaining transcripts and survey responses. The codes were discussed and grouped into themes, after which the team conducted a second round of coding to refine the themes.

\section{RESULTS}
This section presents our study findings, supported by survey data. First, we show the survey results collected from our field studies, along with the quantitative analysis of these results. Next, we describe the value of the mixed-method data collected by Dynamic Surveys for survey stakeholders. Following this, we present how Dynamic Surveys elicits deeper insights from both respondents and stakeholders. Finally, we describe how Dynamic Surveys fosters community insights.

\subsection{Platform Data}

\subsubsection{Overview.}
In our field studies, we collected responses for two surveys, each conducted over a one-week period. For the Career Workshop Survey, a total of 52 participants completed Dynamic Surveys. The platform generated 6 clusters from these responses, with ranking scores ranging from 0 to 68 and an average cluster accuracy of 83.8 (on a 100-point scale) (\textbf{Table \ref{tab:interview1}}). For the Department Course Curriculum Survey, we received responses from 41 participants, resulting in the creation of 9 clusters. These clusters had ranking scores ranging from 18 to 63, with an average cluster accuracy of 92.7(\textbf{Table \ref{tab:interview2}}).

\begin{figure}[b]
  \centering
  \includegraphics[width=1\linewidth]{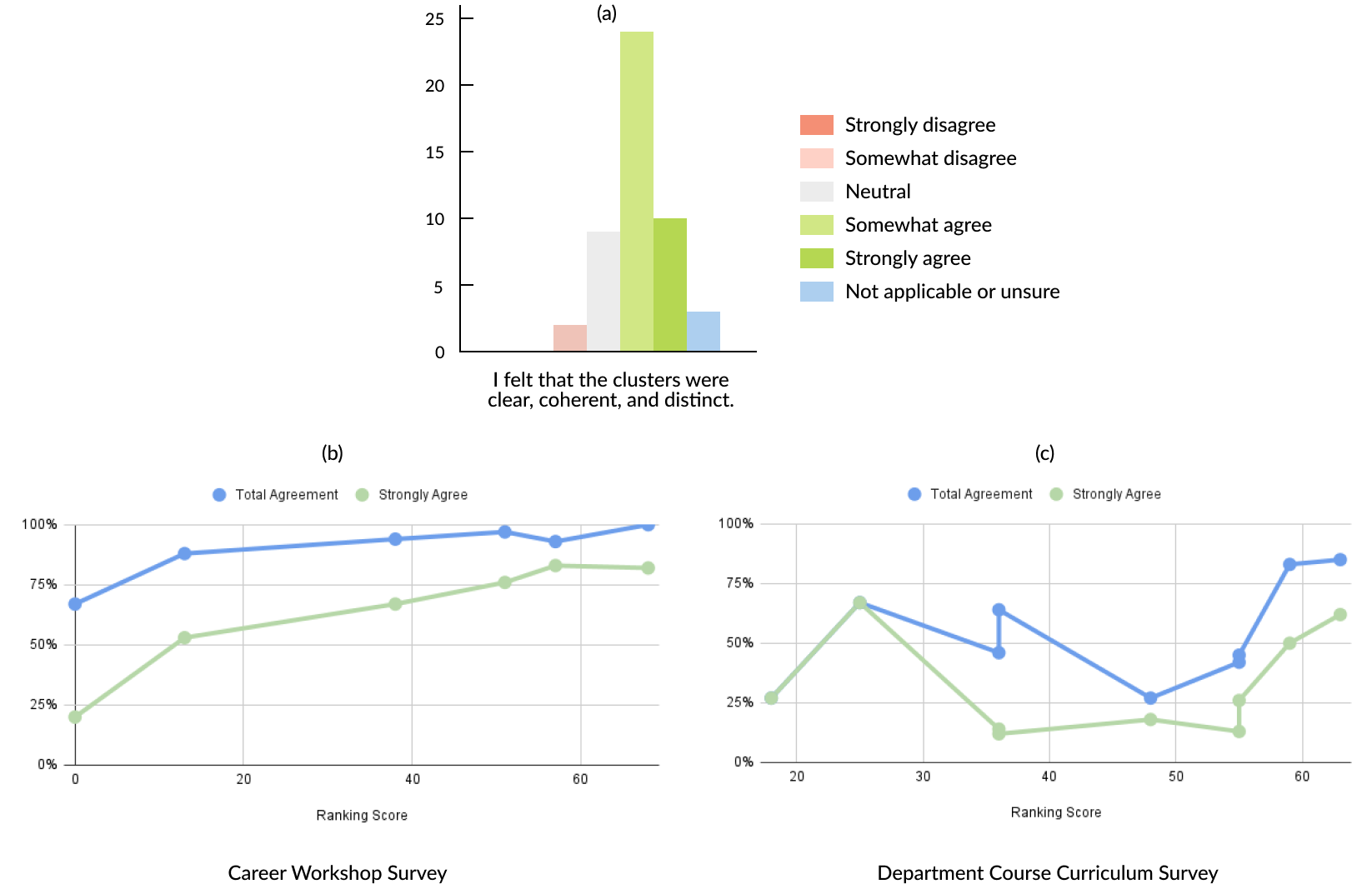}
  \caption{(a) Respondents' answers to the post-study survey question, “I felt that the clusters were clear, coherent, and distinct.” (b) The diagram shows the relationship between each cluster's ranking score the percentage of participants who selected ''agree'' (which includes 'somewhat agree' and ''strongly agree'') as well as the percentage who specifically selected ''strongly agree'' in the Career Workshop Survey. (c) The diagram shows the relationship between each cluster's ranking score the percentage of participants who selected ''agree'' (which includes 'somewhat agree' and ''strongly agree'') as well as the percentage who specifically selected ''strongly agree'' in the Department Course Curriculum Survey.}
  \label{fig:cluster data}
\end{figure}

\begin{table}
\centering
\caption{The clustering results of the Career Workshop Survey.}
\begin{tabular}{p{0.09\textwidth}|p{0.29\textwidth}|p{0.10\textwidth}|p{0.10\textwidth}|p{0.10\textwidth}|p{0.12\textwidth}}
\toprule
Ranking & Cluster Title & Ranking score & Cluster Accuracy & Number of \newline responses & Total \newline Agreement \\
\hline
1 & Interview And Training \newline Processes In Tech & 68 & 83.3 & 9 & 100\% \\
\hline
2 & Qualities Of Top Tech \newline Students & 57 & 93.7 & 24 & 93\% \\
\hline
3 & Strategies For Effective Self Presentation & 51 & 86.0 & 25 & 97\% \\
\hline
4 & Career Pathways And \newline Transitions In Tech & 38 & 90.0 & 5 & 94\% \\
\hline
5 & Work Environment And \newline Culture In Tech Companies & 13 & 100.0 & 1 & 88\% \\
\hline
6 & Intersection Of Policy And Technology & 0 & 50.0 & 2 & 67\% \\
\bottomrule

\end{tabular}
\label{tab:interview1}
\end{table}

\begin{table}
\centering
\caption{The clustering results of the Department Course Curriculum Survey.}
\begin{tabular}{p{0.09\textwidth}|p{0.29\textwidth}|p{0.10\textwidth}|p{0.10\textwidth}|p{0.10\textwidth}|p{0.12\textwidth}}
\toprule
Ranking & Cluster Title & Ranking Score & Cluster Accuracy & Number of \newline responses & Total \newline Agreement \\
\hline
1 & Lack Of Career Preparation Resources In Curriculum & 63 & 90.0 & 10 & 85\% \\
\hline
2 & Need For Job Market \newline Alignment In Courses & 59 & 95.0 & 20 & 83\% \\
\hline
3 & Need For Skills Leveling In Introductory Courses & 55 & 85.7 & 7 & 45\% \\
\hline
4 & Lack Of Interdisciplinary Course Offerings & 55 & 80.0 & 5 & 42\% \\
\hline
5 & Lack Of Prerequisite \newline Knowledge In Courses & 48 & 100.0 & 2 & 27\% \\
\hline
6 & Absence Of Required \newline Networking Courses & 36 & 100.0 & 1 & 64\% \\
\hline
7 & Difficulty Progression In \newline Programming Courses & 36 & 100.0 & 7 & 46\% \\
\hline
8 & Concerns About Discrimination And Bias In Curriculum & 25 & 100.0 & 2 & 67\% \\
\hline
9 & Unreasonable Course \newline Workloads & 18 & 83.3 & 3 & 27\% \\
\bottomrule
\end{tabular}
\label{tab:interview2}
\end{table}

\subsubsection{Cluster quality.}
We evaluated cluster quality by assessing cluster accuracy, which measures how closely responses align with the theme of each cluster (See \textbf{Table \ref{tab:interview1}} and \textbf{Table \ref{tab:interview2}}). Responses within each cluster were manually coded as follows: 1 point for a strong thematic fit, 0.5 points for a weak fit, and 0 points for no fit. 
A response was considered a strong fit if it clearly aligned with the key idea of the cluster summary, and a weak fit if it was only partially relevant or tangentially related.
For instance, in the Career Workshop Survey, \textit{``How well does course material actually translate into the field (e.g., CS projects/lectures vs. CS job)?''} was coded as 0 for the cluster “Interview And Training Processes In Tech,” while \textit{``How to break into entry level roles. Where to find unpaid work in tech.''} was coded as 0.5.
Cluster accuracy was then calculated by dividing the total assigned points by the maximum possible points, resulting in a score on a 0-100 scale. Coding was performed by multiple members of the research team, who independently coded responses and then discussed differences to reach consensus. Across the two surveys, clusters reached an average accuracy of 88.9. Additionally, according to our post-study survey results, 70.8\% of participants found the clusters to be clear, coherent, and distinct (\textbf{Figure \ref{fig:cluster data} (a)}).

\subsubsection{Ratings and rankings.}\label{ratings and rankings}
To assess the value of the rating and ranking results in each survey report, we compared each cluster’s ranking score with the proportion of participants who agreed with each cluster, including those who selected "somewhat agree" and "strongly agree" on the Likert scale (\textbf{Figure \ref{fig:cluster data} (b)} and \textbf{(c)}). In the Career Workshop Survey, we observed a positive correlation between ranking scores and agreement proportions, suggesting that participants' top-ranked topics aligned with their perceived importance. In the Department Course Curriculum Survey, alignment was observed in the top half of clusters but not in the lower-ranked ones. This difference is likely due to the later formation of several clusters in the lower half, which meant fewer participants rated and ranked them, potentially introducing bias.

\subsubsection{Follow-up responses.}
To understand whether responses to follow-up questions could provide new insights into the original questions, we applied the same clustering method to the follow-up responses in both surveys. We found that these follow-up responses revealed additional themes related to the survey topics in both surveys.

In the Career Workshop Survey, the newly identified clusters included:
\begin{itemize}
    \item \textbf{Perception of Competitive Job Market:} Perception of Competitive Job Market: Focuses on feelings of being outmatched by peers in a crowded and competitive job market. 
    \item \textbf{Truthfulness in Resume Submissions for Tech Jobs:} Focuses on honesty and accuracy in resume submissions, specifically regarding applicants' experience and qualifications in the tech industry.
\end{itemize}

In the Department Course Curriculum Survey, the newly identified clusters are the following:
\begin{itemize}
    \item \textbf{Mentorship And Research Guidance Improvement:} Mentorship enhances student support and pathways for engaging in research.
    \item \textbf{Enhancement Of Problem-Solving And Collaboration Skills:} Emphasizes developing problem-solving and teamwork abilities through experiential learning.
\end{itemize}

To further assess the depth provided by follow-up responses compared to their corresponding original responses, we conducted qualitative analysis on the survey results. We found that follow-up responses often added three types of deeper information or insights: (1) further explanations, (2) concrete examples, and (3) suggestions for improvements. Below are examples illustrating each category:
\begin{itemize}
    \item \textbf{Further explanations}
        \begin{itemize}
            \item \textit{Original: }``CSE X is way harder than Y.''
            \item \textit{Follow-up: }``Compared to Y, CSE X often requires students to work more closely with hardware, understanding memory management, pointers, and how the operating system interacts with programs. The complexity of these low-level languages can be tough for students more familiar with higher-level languages like Python or Java.''
        \end{itemize}
    \item \textbf{Concrete examples}
        \begin{itemize}
            \item \textit{Original: }``We have some good courses that match the skills of today’s CS industry, but some courses are outdated. We need more courses matching the industry.''
            \item \textit{Follow-up: }``AI and NLP courses are outdated. We should add topics like ChatGPT and other current established AI models, or topics on creating our own chatbots.''
        \end{itemize}
        \item \textbf{Suggestions for improvements}
        \begin{itemize}
            \item \textit{Original: }``Some courses have an unreasonable workload… making it difficult for students to balance their academic and personal lives… It would be beneficial to adjust the course structure, set appropriate prerequisites, and balance the workload to improve the overall learning experience.''
            \item \textit{Follow-up: }``Adjust core requirements for CS students: Removing non-essential components, like electrical engineering in required classes (e.g., CSE X) for most computer science students, could reduce unnecessary workload.''
        \end{itemize}
\end{itemize}

\subsection{Dynamic Surveys Offer Valuable Quantified Qualitative Insights}
All four survey stakeholders we interviewed recognized the value of the results displayed in the survey report. They noted that, in many scenarios, Dynamic Surveys could provide survey creators with richer insights than traditional survey tools. As one stakeholder shared, 
\begin{quote}
    \textit{``I think especially in a classroom setting or for more academic-style questions, this is probably a far better tool than anything I've seen for gaining a fuller picture of what everyone's thinking.''} (I1)
\end{quote}

\subsubsection{Survey stakeholders found the cluster-based results valuable to them.}
Survey stakeholders noted that Dynamic Surveys offered \textit{``a way of taking open-ended qualitative survey data and imposing some mathematical or quantifiable analysis on it.''} (I4). They liked how the platform provided them with \textit{``clear''} (I3) and \textit{``more quantifiable and structured data''} (I1). They appreciated the cluster-based presentation of data, describing the survey report as \textit{``clean-looking and readable''} (I2) and highlighting the usefulness of \textit{``a good summary section that ... I could look into quickly and be able to see what was going on''} (I1). Participants found value in the rankings and ratings provided for each cluster, which enabled them to \textit{``quickly go through, read the summary, and know what’s top of mind''} (I1). Additionally, the opinion distribution within each cluster \textit{``provides a very unique presentation of data''} (I2), rather than \textit{``just yes or no or something very simple like that''} (I1). Respondents also found the report layout effective, as one participant said, \textit{``The structure of the report, I think, is really excellent... it helps you to understand, you know, what’s coming... and then it goes into a deeper dive into each, so that the structure is great''} (I4).

\subsubsection{Clusters generated by GPT provided meaningful insights but with certain limits for survey stakeholders.}
All survey stakeholders indicated that GPT’s generated clusters exceeded their expectations, though some noted there was still room for improvement. I4 highlighted GPT’s ability to provide more \textit{``objective''} analysis results, as she said, \textit{``It takes the emotion out, it takes the personal perspective out, which I think is really important.''} (I4). She further explained, \textit{··If I'm the person [who codes the data]… I might be in the back of my mind making excuses for the things that people complained about... GPT doesn’t care. It’s not going to get its feelings hurt or feel defensive or angry about this.''} (I4) Additionally, I3, with extensive expertise in qualitative research, was initially skeptical about using GPT for data collection and analysis. However, after viewing the survey report, she said,  \textit{``I really appreciate it's making me appreciate GPT... like we're still even beginning to understand how GPT can help make our jobs richer and better, like this''} (I3). She also pointed out the limitations of the cluster-based results provided by Dynamic Surveys, suggesting that \textit{``understanding how these things are related to each other''} (I3) could provide valuable information in addition to the cluster rankings. Similarly, I2 noted that a few clusters still lacked sufficient distinction for her purposes.

\subsubsection{Dynamic Surveys provided a more efficient way of data collection for non-expert survey creators.}
Dynamic Surveys makes data collection and insight extraction more accessible for survey stakeholders with limited data analysis experience by offering structured reports. For instance, I1, who is not an experienced qualitative researcher, compared Dynamic Surveys to a few other survey tools she had used, noting that those tools provided only basic analytical features, such as \textit{``counting keywords that everyone has used''} (I1), and were \textit{``a little bit clunky''} (I1) to use. Another stakeholder, I4, with limited experience in qualitative research in HCI, noted that she \textit{``could get good at it (Dynamic Surveys), and create really useful surveys in 20 minutes... that’s amazing''} (I4). She also expressed interest in using Dynamic Surveys to \textit{``put together a survey based on that workshop for the participants next week''} (I4). For experienced qualitative researchers like I2 and I3, Dynamic Surveys was also seen as a \textit{``powerful''} (I3) tool for \textit{``people who aren’t trained in analyzing data...because they don't maybe know where to start.''} (I2)

\subsection{Dynamic Surveys Enable Respondents to Share Deeper Insights} \label{deeper}

Alongside providing valuable quantified qualitative data, both respondents and survey stakeholders found that Dynamic Surveys effectively elicited in-depth insights. I3 noted, \textit{``you're getting the richness of an interview or a focus group interview… It's somewhere in between an interview and a survey''} (I3). Similarly, one respondent said, 
\begin{quote}
    \textit{``It allowed me to provide more nuanced responses rather than just sticking to multiple-choice questions. The option to expand on my thoughts added value to the feedback I could give.''} (P16)
\end{quote}

\begin{figure}[b]
  \centering
  \includegraphics[width=0.9\linewidth]{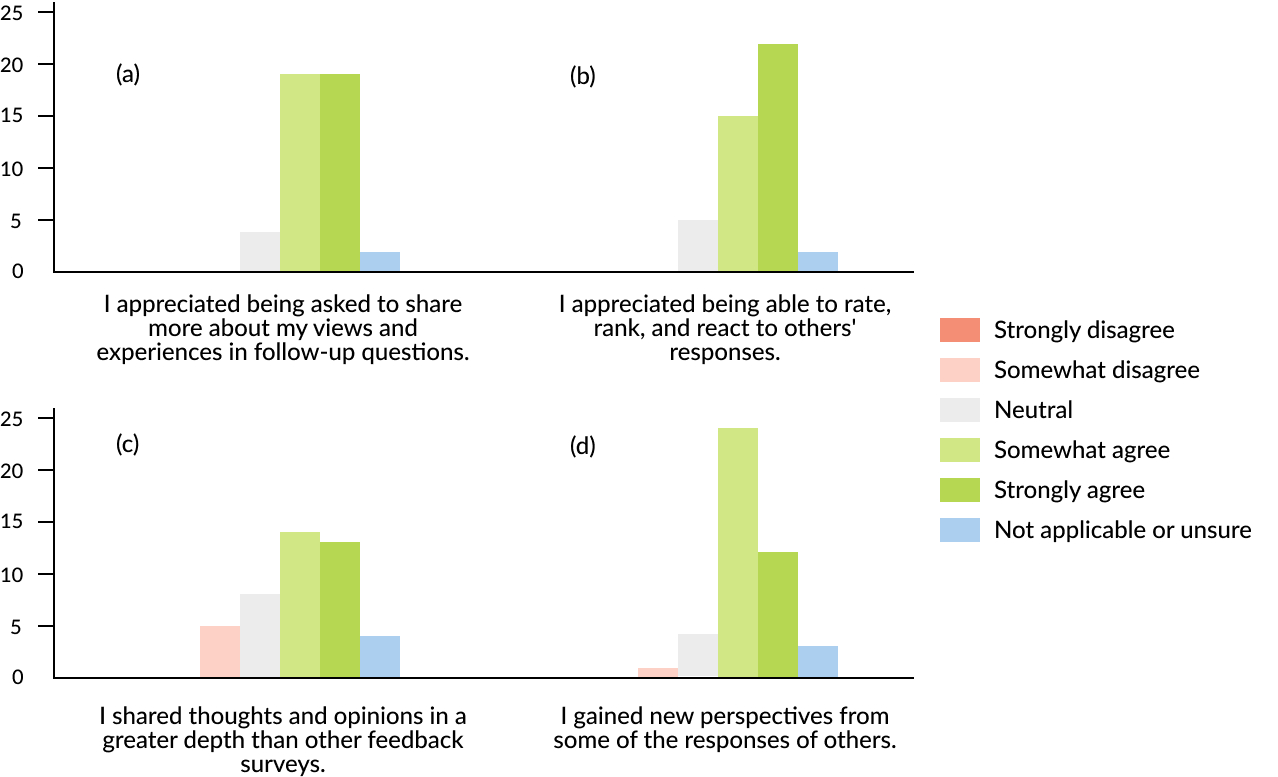}
  \caption{(a) Respondents' answers to the post-study survey question, “I appreciated being asked to share more about my views and experiences in follow-up questions.” (b) I appreciated being able to rate, rank, and react to others' responses.” (c) Respondents' answers to the post-study survey question, “I shared thoughts and opinions in a greater depth than other feedback surveys.” (d) Respondents' answers to the post-study survey question, “I gained new perspectives from some of the responses of others.”}
  \label{fig:post-study survey}
\end{figure}

\subsubsection{Follow-Up questions encourage more thoughtful responses.}
Participants liked the follow-up question feature, as \textit{``it can help the creators understand my line of thinking more''} (P28) (\textbf{Figure \ref{fig:post-study survey} (a)} and \textbf{(c)}). In contrast to the more general questions in other survey tools, many respondents felt that Dynamic Surveys offered more tailored and \textit{``specific''} follow-up questions, which enabled them to \textit{``better  describe my thoughts and opinions regarding the topics''} (P29). 
This \textit{``personalized''} experience was a key reason for their preference for Dynamic Surveys. 

In addition, survey stakeholders also found value in the follow-up questions, noting that \textit{``the follow-up response is a much richer response than the first [response to the original question]''} (I3) and that \textit{``this element of feedback can be very motivating''} (I2) for respondents. Additionally, respondents also saw potential benefits in using Dynamic Surveys as survey creators themselves, as one said,
\begin{quote}
    \textit{``For the class I’m TAing, which focuses on developing games in Twine, I gathered feedback from students on what aspects of Twine they want to explore further. The responses I received varied—some were quite specific, while others were broad. I think this tool could help me follow up with those who provided more general ideas.''} (P27)
\end{quote}
\subsubsection{Comparing rankings with others promotes self-reflection.}\label{reflection}
Respondents appreciated the chance to compare their rankings with others, encouraging reflection on their views relative to broader trends (\textbf{Figure \ref{fig:resonance} (b)} and \textbf{(d)}). One participant noted that \textit{``It was interesting to see how my values align with the peers in my group and how they do not. It makes me reconsider the options and evaluate my values on a deeper level.''} (P8) Respondents felt this comparison feature made Dynamic Surveys \textit{``not just about me clicking and submitting; it actually makes me reflect on my responses.''} (P8) and even \textit{``prompted me to think more critically about my responses''} (P5).

\subsubsection{While Dynamic Surveys offer in-depth insights, they lack the depth of other qualitative methods.}

Both I2 and I3, experienced qualitative researchers, viewed Dynamic Surveys as \textit{``unique and helpful''} (I2), yet still noted that it \textit{``lacks certain interactive depth compared to qualitative interviews''} (I2). I3 found value in the feature that allows respondents to compare rankings and provide explanations, as it mirrors her focus group experience, where \textit{``people will say something, and it’ll give somebody else an idea, and so it will lead to sort of a richer conversation.''} (I3). However, she also mentioned that she prefers to \textit{``phrase questions in a way that encourages storytelling rather than asking respondents to condense their whole experience''} (I3) to \textit{``get richer data.''} (I3) She pointed out that Dynamic Surveys offer limited opportunities for this kind of interaction, which is why she often prefers other methods over surveys, as she has found herself \textit{``not happy with self-report data.''} (I3) Similarly, while I2 appreciated that surveys \textit{``are good in a way because respondents might not get as defensive as in interviews''} (I2), she noted that in interviews, \textit{``there’s something about being face-to-face or in-depth that encourages participants to share more''} (I2).

\subsection{Dynamic Surveys Increase Engagement and Foster a Sense of Community}\label{engagement}
One distinctive feature of Dynamic Surveys, compared to other survey tools, is that participants can view others' responses anonymously. Our findings show that participants found this experience both novel and engaging, which enhanced their sense of involvement.

\subsubsection{Follow-up questions and the ranking system make respondents feel engaged and involved.}

The personalized open-ended questions and the ability to view others’ responses made participants feel more engaged with Dynamic Surveys. Several participants mentioned that the tailored follow-up questions \textit{``made me feel like the survey was listening to me''} (P10). Participants also appreciated how Dynamic Surveys presented others' answers, as one said, \textit{``I liked it a lot, especially the feedback since I don't typically engage with the results of surveys I take''} (P18). Others echoed this sentiment and found it \textit{``interesting to see how my values align with the peers in my group and how they do not.''} (P8) They felt that \textit{``seeing how others interpret the same things makes me feel more involved''} (P2) and valued the chance \textit{``to rate and react to others’ responses for deeper engagement and gaining new perspectives''} (P44).

\begin{figure}[b]
  \centering
  \includegraphics[width=0.8\linewidth]{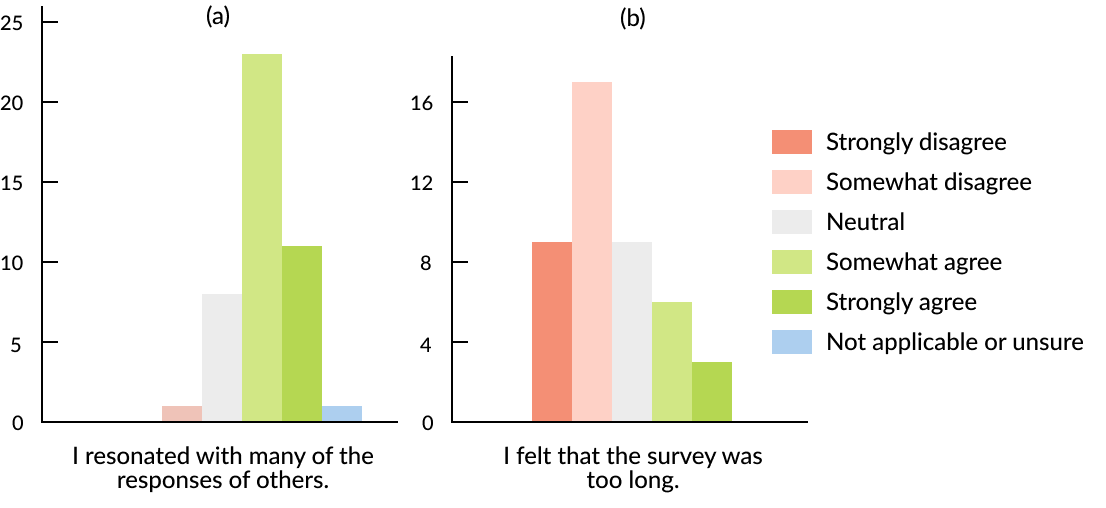}
  \caption{(a) Respondents' answers to the post-study survey question, “I resonated with many of the responses of others.” (b) Respondents' answers to the post-study survey question, “I felt that the survey was too long.”}
  \label{fig:resonance}
\end{figure}

\subsubsection{Dynamic surveys fostered community connection and collaborative engagement through shared responses.} \label{community}
Participants \textit{``liked being able to see and rank other people's ideas, because I found ideas that I really resonated with''} (P25), and it \textit{``assured me I share similar concerns with others and allowed me to know what others are thinking''} (P14) (\textbf{Figure \ref{fig:resonance} (a)}). Additionally, some felt more connected to others in the community, as one noted, \textit{``Being able to see other people's responses in an organized format makes me feel closer to the needs and feelings of my peers''} (P33). Rather than simply answering questions individually, participants felt Dynamic Surveys provided \textit{``a really great way to share how others within the community feel about these different topics''} (P22) and \textit{``made me reflect on how my views contributed to the overall results''} (P8). Importantly, Dynamic Surveys made respondents feel \textit{``collaborative''} by allowing them to \textit{``give my opinion on other people's responses… and mix the responses together into a more cohesive answer to the overall question''} (P13).

\subsubsection{While people feel engaged, some found the survey too long.} \label{lengthiness}

While respondents liked the personalized experience of Dynamic Surveys, a few found the surveys lengthy (\textbf{Figure \ref{fig:resonance} (b)}). One major factor was the increase in clusters, which led to feeling overwhelmed by the rating and ranking tasks, particularly when there were \textit{``10+ options to rank amongst''} (P44). One noted, \textit{``As much as I liked the clusters, I did feel like there might be some overload on the eyes''} (P3). Additionally, the number of questions felt burdensome for some, making it harder for them to stay motivated to complete the survey, as one explained, \textit{``I felt like there were a lot of open-ended questions, which made the process a bit tiring''} (P12). One participant mentioned that they initially expected to answer only the original open-ended question but then realized that \textit{``more questions were added as I was completing it so it felt a bit frustrating''} (P8).

\section{DISCUSSION}
Our findings revealed that Dynamic Surveys provided valuable quantitative insights for survey creators, prompted deeper responses, and fostered connections within the community. We also reported the limitations found by participants in using Dynamic Surveys. In this section, we explore potential improvements for data accuracy in Dynamic Surveys and suggest two design directions: centering LLM-powered personalization in survey design and developing community-based surveys to encourage collaboration among respondents.

\subsection{Advancing Personalized Mixed-Method Survey Design through LLMs} \label{LLM discussion}

This study shows that personalized survey design enhances respondent engagement and encourages deeper insights, helping survey creators better understand respondents' perspectives and reasoning (\textbf{Section \ref{deeper}} and \textbf{Section \ref{engagement}}). By leveraging the advanced language-processing capabilities of LLMs, it enables survey to provide more adaptive and smart questions that can adjust dynamically to capture subtle nuances in participants' responses. This personalization supports a more efficient and insightful data collection process.

Personalized questions centered on individual responses enable researchers to gain deeper insights. In this study, follow-up questions were generated based on single responses, but future designs could consider questions that draw connections across multiple responses from a participant, approaching a level of comprehension similar to that of a human qualitative researcher.

Moreover, LLM-based questions have the potential to expand the diversity of insights by drawing in a broader range of perspectives from respondents, as seen in Dynamic Surveys, where respondents interact with others' rankings. By designing personalized matching mechanisms, LLMs could tailor questions by pairing respondents with specific themes or perspectives based on their previous opinions,  encouraging them to contribute diverse viewpoints to enrich data themes. 

That is to say, personalized survey design could not only offer tailored questions for respondents but also address the varied needs of survey creators. For instance, survey stakeholders in this study, such as I3, valued descriptive, story-driven responses. In such cases, follow-up questions could focus on eliciting more detailed personal experiences. In contrast, surveys intended for online deliberation might emphasize collective consensus and divergence on the survey topic, exploring potential connections within group responses.

In addition, future LLM-driven survey designs could also enhance the personalization experience through multidimensional data. While some research has explored data-driven survey designs, these surveys typically rely on raw, unprocessed data. Given LLMs' ability to interpret data, they could be applied to various data types beyond text and support diverse analytical techniques, such as sentiment analysis for voice data, to create a more comprehensive survey experience. For instance, in mental health research, follow-up questions could be adapted based on respondents' contextual and emotional cues to gather more valuable data.

Our findings identify a key challenge in balancing question depth with respondent burden. Personalized follow-up questions can increase survey length, potentially overwhelming respondents and affecting response rates and data quality. Prior work has shown that the framing of survey instructions can influence participants’ perceived burden~\cite{yu2015can}. Building on this, future designs could explore how LLMs might tailor the wording of follow-up questions not only to elicit deeper insights but also to reinforce the value of participants’ contributions, which may enhance engagement and reduce fatigue. In addition, as the number of clusters increases during the rating phase of Dynamic Surveys (\textbf{Section \ref{lengthiness}}), the repetitive format of the questions might lead to respondent fatigue or disengagement~\cite{beatty2019advances}. Future iterations could explore presenting clusters in smaller sets or across multiple pages to sustain attention and reduce cognitive load. Moreover, LLMs could serve as facilitators in surveys, interpreting real-time feedback to monitor engagement levels. When indicators of respondent strain, such as shorter or repetitive answers, are detected, LLMs could dynamically adjust by simplifying subsequent questions or moderating depth, helping to sustain respondents’ attention and maintain data quality.

Finally, the use of LLMs in survey workflows raises important ethical considerations. Because responses are transmitted to third-party APIs for processing, researchers must consider risks related to data privacy and external data handling. While our study focused on minimal-risk topics, surveys involving sensitive information demand greater transparency about how data is used and stored, and stronger safeguards to ensure informed consent. Additionally, while LLMs can support personalized and adaptive survey experiences at scale, their outputs may still reflect biases or miss contextual nuances in participants’ input. In such settings, human oversight remains essential to ensure that generated clusters and follow-up questions are appropriate, meaningful, and aligned with research goals.

\subsection{Designing Collaborative Surveys for Reflection and Collective Expression}
While common survey tools don’t support respondents in viewing survey results, our study suggests that offering respondents the opportunity to see and engage with others’ responses could benefit both individuals and the community. It not only prompts personal reflection on the survey topic (\textbf{Section \ref{reflection}}) but also fosters community-centered connection (\textbf{Section \ref{community}}), allowing respondents to build on each other’s perspectives and develop shared insights. Such interaction can enrich the collected data, providing survey creators with more nuanced and meaningful insights.

Collaborative surveys are not a novel concept in the literature. They have been explored for a long time in participatory democracy, where surveys are used to support collective decision-making to more accurately reflect the community’s needs. Our work is inspired by this research and, as mentioned earlier, by Wikisurveys~\cite{salganik2015wiki}, which highlights a unique aspect of collaborative surveys where each respondent contributes not as an isolated feedback provider but adds value to the community's shared understanding by building on others' opinions. Extending this concept to broader survey contexts could enable survey creators to gather continuously evolving datasets that reflect diverse perspectives solely from the respondents' viewpoints, reducing the influence of researcher bias. 

Our work represented a step toward creating surveys that blend both qualitative and quantitative data collection to support collaborative participation. Although Dynamic Surveys enables participants to express their views on ranking differences, its collaborative interactions remain limited. Providing respondents with an additional opportunity to share insights after reviewing survey results with detailed data may lead to more deeper insights. Future work holds great potential for designing surveys that foster meaningful collaboration in diverse formats, perhaps by integrating personalized mechanisms discussed in Section 6.1 to support more interactive conversations, such as through role-based matching and multi-round information exchanges.

Finally, further research could explore how collaborative survey methods impact community building. One potential direction could be integrating these surveys with other activities to foster a more connected and engaged community.

While this study did not aim to evaluate consensus-building, we see promising opportunities for future research to explore how the collaborative dynamics of Dynamic Surveys might be applied in more deliberative contexts. In settings such as participatory budgeting, community planning, or stakeholder engagement, researchers are often not only interested in gathering diverse perspectives but also in understanding how differing views can be negotiated or prioritized~\cite{elwood1998gis, schugurensky2024participatory, okada2013community}. Although our current work focuses on eliciting a broad range of insights rather than fostering agreement, future adaptations of this approach may support processes where interaction among participants contributes to shared understanding or collective decision-making. In particular, community-based research often requires both the collection of input and the facilitation of meaningful dialogue~\cite{mcdavitt2016dissemination, montoya2011dialogical}. Designing survey tools that integrate these functions could help surface points of convergence and divergence in a way that supports both reflection and collective sense-making. Exploring this intersection represents a promising direction for extending the methodological potential of Dynamic Surveys.

\subsection{Designing User Interfaces to Support Interpretation and Survey Creation}
In \textbf{Section \ref{ratings and rankings}} we noted that the sequence in which clusters are generated and the number of participants rating and ranking each cluster may affect the accuracy of the quantitative data. If participants are presented with clusters created at different times, it may also influence their ratings and rankings, potentially resulting in noticeable discrepancies between later respondents' responses and earlier cumulative results. However, survey creators might lack the abilities to assess the reliability of these quantitative data. As a result, user interfaces that support data monitoring and quality control are needed. A possible design solution is to include a dashboard within the survey report that offers real-time statistics on cluster ratings, such as standard deviations to show how much ratings have converged, helping creators estimate the additional responses needed for more reliable data. Additionally, future work could focus on refining clustering methods to further improve the accuracy of quantitative results.

In addition to improving the interpretability of cluster-based results, future research could explore ways to support survey creators—especially those without technical backgrounds—in anticipating how their inputs may influence the downstream interaction. While the survey creation process is intentionally lightweight, requiring only a single open-ended question, some users may find it difficult to envision how participants will engage with the system or how their prompt may shape the flow of follow-up questions. One possible direction is to develop interfaces that simulate sample responses and show how these would be clustered and reflected back to participants. For example, creators could preview how a prompt might lead to particular follow-up phrasing or themes, helping them adjust their wording to better align with their intent. This type of preview-based feedback could enhance creators’ confidence in using the system and broaden its accessibility across domains where formal research training is uncommon.

\section{LIMITATIONS}
Several limitations of our study should be acknowledged. First, our survey platform currently focuses on a single topic per survey, which may not fully meet the needs of all survey creators. On Dynamic Surveys, each survey is designed around one original open-ended question. If we enable survey creators to add more core questions, might significantly increase the survey's length and reduce response rates. Although it limits the diversity of topics covered in Dynamic Surveys, it can still be used in various contexts to collect data, such as classroom feedback, collective decision-making, and crowdsourcing ideas. Future work could explore varied survey formats to address a broader range of creator needs. Second, our study was conducted in two contexts, on a campus within a narrow demographic group, and with a limited number of survey stakeholders, which also constrained the number and diversity of survey stakeholders we were able to recruit. While the stakeholders we interviewed had different qualitative research backgrounds, offering valuable insights, this limited sample reflects the constraints of our study contexts rather than the full range of potential users. Additional studies in more varied settings could further help us identify the potential of LLM-based dynamic surveys and highlight new opportunities and challenges. Third, our survey design might introduce certain biases. For example, asking whether clusters are “clear, coherent, and distinct” conflates multiple dimensions, which might make participant responses harder to interpret. Additionally, our use of agreement-based scales could be subject to acquiescence bias, especially since respondents were evaluating a system developed by the researchers. We acknowledge these limitations and suggest that future work explore alternative wording and scale formats to reduce such effects.

\section{CONCLUSION}
In this paper, we explored the blend of qualitative and quantitative data analysis to create a dynamic and interactive survey experience. We presented our design and development of Dynamic Surveys, a platform that leverages LLMs for real-time clustering and allows respondents to rate, rank, and engage with peers’ feedback. To evaluate Dynamic Surveys, we conducted two real-world studies involving 97 participants across two settings to understand their user experience and views of the data presented in the survey reports. Our findings indicated that the platform provided valuable, quantifiable qualitative data and deeper insights for survey stakeholders, while also increasing engagement and fostering a sense of community. We concluded by discussing design implications for creating LLM-based survey tools that integrate qualitative, quantitative and community-centered data collection for future work.

\section{Acknowledgments}
We would like to thank the anonymous CSCW reviewers for their helpful feedback, which helped us significantly improve this paper. We are also grateful to the survey stakeholders who collaborated with us on this project and provided valuable feedback. Finally, we thank the many undergraduate students in the Tech4Good Collaboratory skill-building teams who contributed to development-related work, including Raghavendra Raikar, José Chavez, Jackson Tran, Reina Itakura, Brandon Ng, Khin Saw Yone, Christine Lee, Jeffrey Chao, Junwei Lu, Ingrid Morgan, and Hari Varsha. This work was supported in part by funding from the Google Research Scholar Program.

\bibliographystyle{ACM-Reference-Format}
\bibliography{paper}

@article{buchanan2009online,
  title={Online survey tools: Ethical and methodological concerns of human research ethics committees},
  author={Buchanan, Elizabeth A and Hvizdak, Erin E},
  journal={Journal of empirical research on human research ethics},
  volume={4},
  number={2},
  pages={37--48},
  year={2009},
  publisher={SAGE Publications Sage CA: Los Angeles, CA}
}

@article{wright2005researching,
  title={Researching Internet-based populations: Advantages and disadvantages of online survey research, online questionnaire authoring software packages, and web survey services},
  author={Wright, Kevin B},
  journal={Journal of computer-mediated communication},
  volume={10},
  number={3},
  pages={JCMC1034},
  year={2005},
  publisher={Oxford University Press Oxford, UK}
}

@article{parker1992collecting,
  title={Collecting data the e-mail way.},
  author={Parker, Lorraine},
  journal={Training \& Development},
  volume={46},
  number={7},
  pages={52--55},
  year={1992},
  publisher={Association for Talent Development (ATD)}
}

@article{mehta1995comparing,
  title={Comparing response rates and response content in mail versus electronic mail surveys},
  author={Mehta, Raj and Sivadas, Eugene},
  journal={Market Research Society. Journal.},
  volume={37},
  number={4},
  pages={1--12},
  year={1995},
  publisher={SAGE Publications Sage UK: London, England}
}

@article{rouder2021all,
  title={What to do with all those open-ended responses? Data visualization techniques for survey researchers},
  author={Rouder, Jessie and Saucier, Olivia and Kinder, Rachel and Jans, Matt},
  journal={Survey Practice},
  year={2021}
}

@inproceedings{ferrario2022eliciting,
  title={Eliciting people’s first-order concerns: Text analysis of open-ended survey questions},
  author={Ferrario, Beatrice and Stantcheva, Stefanie},
  booktitle={AEA Papers and Proceedings},
  volume={112},
  pages={163--169},
  year={2022},
  organization={American Economic Association 2014 Broadway, Suite 305, Nashville, TN 37203}
}

@article{denzin1996handbook,
  title={Handbook of qualitative research},
  author={Denzin, Norman K and Lincoln, Yvonna S},
  journal={Journal of Leisure Research},
  volume={28},
  number={2},
  pages={132},
  year={1996},
  publisher={National Recreation and Park Association}
}

@book{maxwell2013qualitative,
  title={Qualitative research design: An interactive approach: An interactive approach},
  author={Maxwell, Joseph A},
  year={2013},
  publisher={sage}
}

@inproceedings{freitas2018case,
  title={In case of doubt see the manual: A comparative analysis of (self) learning packages qualitative research software},
  author={Freitas, F{\'a}bio and Ribeiro, Jaime and Brand{\~a}o, Catarina and de Souza, Francisl{\^e} Neri and Costa, Ant{\'o}nio Pedro and Reis, Lu{\'\i}s Paulo},
  booktitle={Computer supported qualitative research: Second international symposium on qualitative research (ISQR 2017)},
  pages={176--192},
  year={2018},
  organization={Springer}
}

@article{wesslen2018computer,
  title={Computer-assisted text analysis for social science: Topic models and beyond},
  author={Wesslen, Ryan},
  journal={arXiv preprint arXiv:1803.11045},
  year={2018}
}

@article{andreotta2019analyzing,
  title={Analyzing social media data: A mixed-methods framework combining computational and qualitative text analysis},
  author={Andreotta, Matthew and Nugroho, Robertus and Hurlstone, Mark J and Boschetti, Fabio and Farrell, Simon and Walker, Iain and Paris, Cecile},
  journal={Behavior research methods},
  volume={51},
  pages={1766--1781},
  year={2019},
  publisher={Springer}
}

@inproceedings{bakharia2016interactive,
  title={Interactive topic modeling for aiding qualitative content analysis},
  author={Bakharia, Aneesha and Bruza, Peter and Watters, Jim and Narayan, Bhuva and Sitbon, Laurianne},
  booktitle={Proceedings of the 2016 ACM on Conference on Human Information Interaction and Retrieval},
  pages={213--222},
  year={2016}
}

@article{lennon2021developing,
  title={Developing and testing an automated qualitative assistant (AQUA) to support qualitative analysis},
  author={Lennon, Robert P and Fraleigh, Robbie and Van Scoy, Lauren J and Keshaviah, Aparna and Hu, Xindi C and Snyder, Bethany L and Miller, Erin L and Calo, William A and Zgierska, Aleksandra E and Griffin, Christopher},
  journal={Family medicine and community health},
  volume={9},
  number={Suppl 1},
  year={2021},
  publisher={BMJ Publishing Group}
}

@inproceedings{stenetorp2012brat,
  title={BRAT: a web-based tool for NLP-assisted text annotation},
  author={Stenetorp, Pontus and Pyysalo, Sampo and Topi{\'c}, Goran and Ohta, Tomoko and Ananiadou, Sophia and Tsujii, Jun’ichi},
  booktitle={Proceedings of the Demonstrations at the 13th Conference of the European Chapter of the Association for Computational Linguistics},
  pages={102--107},
  year={2012}
}

@article{crowston2010machine,
  title={Machine learning and rule-based automated coding of qualitative data},
  author={Crowston, Kevin and Liu, Xiaozhong and Allen, Eileen E},
  journal={proceedings of the American Society for Information Science and Technology},
  volume={47},
  number={1},
  pages={1--2},
  year={2010},
  publisher={Wiley Online Library}
}

@article{li2021qualitative,
  title={Qualitative coding in the computational era: A hybrid approach to improve reliability and reduce effort for coding ethnographic interviews},
  author={Li, Zhuofan and Dohan, Daniel and Abramson, Corey M},
  journal={Socius},
  volume={7},
  pages={23780231211062345},
  year={2021},
  publisher={SAGE Publications Sage CA: Los Angeles, CA}
}

@article{guetterman2018augmenting,
  title={Augmenting qualitative text analysis with natural language processing: methodological study},
  author={Guetterman, Timothy C and Chang, Tammy and DeJonckheere, Melissa and Basu, Tanmay and Scruggs, Elizabeth and Vydiswaran, VG Vinod},
  journal={Journal of medical Internet research},
  volume={20},
  number={6},
  pages={e231},
  year={2018},
  publisher={JMIR Publications Toronto, Canada}
}

@inproceedings{lejeune2011normal,
  title={From normal business to financial crisis... and back again. An illustration of the benefits of Cassandre for qualitative analysis},
  author={Lejeune, Christophe},
  booktitle={Forum: Qualitative Sozialforschung},
  volume={12},
  number={1},
  year={2011},
  organization={Institut fur Klinische Sychologie and Gemeindesychologie, Germany}
}

@article{yan2014semi,
  title={Semi-automatic content analysis of qualitative data},
  author={Yan, Jasy Liew Suet and McCracken, Nancy and Crowston, Kevin},
  journal={IConference 2014 Proceedings},
  year={2014},
  publisher={iSchools}
}

@inproceedings{marathe2018semi,
  title={Semi-automated coding for qualitative research: A user-centered inquiry and initial prototypes},
  author={Marathe, Megh and Toyama, Kentaro},
  booktitle={Proceedings of the 2018 CHI conference on human factors in computing systems},
  pages={1--12},
  year={2018}
}

@article{bazeley2009analysing,
  title={Analysing qualitative data: More than ‘identifying themes’},
  author={Bazeley, Pat},
  journal={Malaysian journal of qualitative research},
  volume={2},
  number={2},
  pages={6--22},
  year={2009},
  publisher={Citeseer}
}

@article{grimmer2013text,
  title={Text as data: The promise and pitfalls of automatic content analysis methods for political texts},
  author={Grimmer, Justin and Stewart, Brandon M},
  journal={Political analysis},
  volume={21},
  number={3},
  pages={267--297},
  year={2013},
  publisher={Cambridge University Press}
}

@article{chen2018using,
  title={Using machine learning to support qualitative coding in social science: Shifting the focus to ambiguity},
  author={Chen, Nan-Chen and Drouhard, Margaret and Kocielnik, Rafal and Suh, Jina and Aragon, Cecilia R},
  journal={ACM Transactions on Interactive Intelligent Systems (TiiS)},
  volume={8},
  number={2},
  pages={1--20},
  year={2018},
  publisher={ACM New York, NY, USA}
}

@article{feuston2021putting,
  title={Putting tools in their place: The role of time and perspective in human-AI collaboration for qualitative analysis},
  author={Feuston, Jessica L and Brubaker, Jed R},
  journal={Proceedings of the ACM on Human-Computer Interaction},
  volume={5},
  number={CSCW2},
  pages={1--25},
  year={2021},
  publisher={ACM New York, NY, USA}
}

@inproceedings{chen2016challenges,
  title={Challenges of applying machine learning to qualitative coding},
  author={Chen, Nan-chen and Kocielnik, Rafal and Drouhard, Margaret and Pe{\~n}a-Araya, Vanessa and Suh, Jina and Cen, Keting and Zheng, Xiangyi and Aragon, Cecilia R and Pe{\~n}a-Araya, V},
  booktitle={ACM SIGCHI Workshop on Human-Centered Machine Learning},
  year={2016}
}

@inproceedings{klie2018inception,
  title={The inception platform: Machine-assisted and knowledge-oriented interactive annotation},
  author={Klie, Jan-Christoph and Bugert, Michael and Boullosa, Beto and De Castilho, Richard Eckart and Gurevych, Iryna},
  booktitle={Proceedings of the 27th international conference on computational linguistics: System demonstrations},
  pages={5--9},
  year={2018}
}

@article{jiang2021supporting,
  title={Supporting serendipity: Opportunities and challenges for Human-AI Collaboration in qualitative analysis},
  author={Jiang, Jialun Aaron and Wade, Kandrea and Fiesler, Casey and Brubaker, Jed R},
  journal={Proceedings of the ACM on Human-Computer Interaction},
  volume={5},
  number={CSCW1},
  pages={1--23},
  year={2021},
  publisher={ACM New York, NY, USA}
}

@inproceedings{rietz2021cody,
  title={Cody: An AI-based system to semi-automate coding for qualitative research},
  author={Rietz, Tim and Maedche, Alexander},
  booktitle={Proceedings of the 2021 CHI conference on human factors in computing systems},
  pages={1--14},
  year={2021}
}

@inproceedings{gebreegziabher2023patat,
  title={Patat: Human-ai collaborative qualitative coding with explainable interactive rule synthesis},
  author={Gebreegziabher, Simret Araya and Zhang, Zheng and Tang, Xiaohang and Meng, Yihao and Glassman, Elena L and Li, Toby Jia-Jun},
  booktitle={Proceedings of the 2023 CHI Conference on Human Factors in Computing Systems},
  pages={1--19},
  year={2023}
}

@inproceedings{hong2022scholastic,
  title={Scholastic: Graphical human-AI collaboration for inductive and interpretive text analysis},
  author={Hong, Matt-Heun and Marsh, Lauren A and Feuston, Jessica L and Ruppert, Janet and Brubaker, Jed R and Szafir, Danielle Albers},
  booktitle={Proceedings of the 35th Annual ACM Symposium on User Interface Software and Technology},
  pages={1--12},
  year={2022}
}

@inproceedings{goldman2022quad,
  title={Quad: Deep-learning assisted qualitative data analysis with affinity diagrams},
  author={Goldman, Ariel and Espinosa, Cindy and Patel, Shivani and Cavuoti, Francesca and Chen, Jade and Cheng, Alexandra and Meng, Sabrina and Patil, Aditi and Chilton, Lydia B and Morrison-Smith, Sarah},
  booktitle={CHI Conference on Human Factors in Computing Systems Extended Abstracts},
  pages={1--7},
  year={2022}
}

@article{zhang2023redefining,
  title={Redefining qualitative analysis in the AI era: Utilizing ChatGPT for efficient thematic analysis},
  author={Zhang, He and Wu, Chuhao and Xie, Jingyi and Lyu, Yao and Cai, Jie and Carroll, John M},
  journal={arXiv preprint arXiv:2309.10771},
  year={2023}
}

@inproceedings{gao2024collabcoder,
  title={CollabCoder: a lower-barrier, rigorous workflow for inductive collaborative qualitative analysis with large language models},
  author={Gao, Jie and Guo, Yuchen and Lim, Gionnieve and Zhang, Tianqin and Zhang, Zheng and Li, Toby Jia-Jun and Perrault, Simon Tangi},
  booktitle={Proceedings of the CHI Conference on Human Factors in Computing Systems},
  pages={1--29},
  year={2024}
}

@article{gao2023coaicoder,
  title={CoAIcoder: Examining the effectiveness of AI-assisted human-to-human collaboration in qualitative analysis},
  author={Gao, Jie and Choo, Kenny Tsu Wei and Cao, Junming and Lee, Roy Ka-Wei and Perrault, Simon},
  journal={ACM Transactions on Computer-Human Interaction},
  volume={31},
  number={1},
  pages={1--38},
  year={2023},
  publisher={ACM New York, NY}
}

@article{zhang2024qualitative,
  title={When Qualitative Research Meets Large Language Model: Exploring the Potential of QualiGPT as a Tool for Qualitative Coding},
  author={Zhang, He and Wu, Chuhao and Xie, Jingyi and Rubino, Fiona and Graver, Sydney and Kim, ChanMin and Carroll, John M and Cai, Jie},
  journal={arXiv preprint arXiv:2407.14925},
  year={2024}
}

@article{khan2024automating,
  title={Automating Thematic Analysis: How LLMs Analyse Controversial Topics},
  author={Khan, Awais Hameed and Kegalle, Hiruni and D'Silva, Rhea and Watt, Ned and Whelan-Shamy, Daniel and Ghahremanlou, Lida and Magee, Liam},
  journal={arXiv preprint arXiv:2405.06919},
  year={2024}
}

@article{booker2021survey,
  title={Survey strategies to increase participant response rates in primary care research studies},
  author={Booker, Quiera S and Austin, Jessica D and Balasubramanian, Bijal A},
  journal={Family Practice},
  volume={38},
  number={5},
  pages={699--702},
  year={2021},
  publisher={Oxford University Press UK}
}

@article{munoz2010improving,
  title={Improving the response rate and quality in Web-based surveys through the personalization and frequency of reminder mailings},
  author={Mu{\~n}oz-Leiva, Francisco and S{\'a}nchez-Fern{\'a}ndez, Juan and Montoro-R{\'\i}os, Francisco and Ib{\'a}{\~n}ez-Zapata, Jos{\'e} {\'A}ngel},
  journal={Quality \& Quantity},
  volume={44},
  pages={1037--1052},
  year={2010},
  publisher={Springer}
}

@article{griggs2018impact,
  title={The impact of greeting personalization on prevalence estimates in a survey of sexual assault victimization},
  author={Griggs, Ashley K and Berzofsky, Marcus E and Shook-Sa, Bonnie E and Lindquist, Christine H and Enders, Kimberly P and Krebs, Christopher P and Planty, Michael and Langton, Lynn},
  journal={Public Opinion Quarterly},
  volume={82},
  number={2},
  pages={366--378},
  year={2018},
  publisher={Oxford University Press US}
}

@article{thomas2024methodological,
  title={Methodological and practical guidance for designing and conducting online qualitative surveys in public health},
  author={Thomas, Samantha L and Pitt, Hannah and McCarthy, Simone and Arnot, Grace and Hennessy, Marita},
  journal={Health Promotion International},
  volume={39},
  number={3},
  pages={daae061},
  year={2024},
  publisher={Oxford University Press US}
}

@article{heritier2023food,
  title={Food \& You: A digital cohort on personalized nutrition},
  author={H{\'e}ritier, Harris and All{\'e}mann, Chlo{\'e} and Balakiriev, Oleksandr and Boulanger, Victor and Carroll, Sean F and Froidevaux, No{\'e} and Hugon, Germain and Jaquet, Yannis and Kebaili, Djilani and Riccardi, Sandra and others},
  journal={PLOS Digital Health},
  volume={2},
  number={11},
  pages={e0000389},
  year={2023},
  publisher={Public Library of Science San Francisco, CA USA}
}

@article{huguenin2017predictive,
  title={A predictive model for user motivation and utility implications of privacy-protection mechanisms in location check-ins},
  author={Huguenin, K{\'e}vin and Bilogrevic, Igor and Machado, Joana Soares and Mihaila, Stefan and Shokri, Reza and Dacosta, Italo and Hubaux, Jean-Pierre},
  journal={IEEE Transactions on Mobile Computing},
  volume={17},
  number={4},
  pages={760--774},
  year={2017},
  publisher={IEEE}
}

@inproceedings{hannah2021interactive,
  title={Interactive survey design using pidgin and GIFS},
  author={Hannah Ajilore, Oluwatoyin and Eloho Malaka, Lauretta and Busayo Sakpere, Aderonke and Grace Oluwadebi, Ayomiposi},
  booktitle={Proceedings of the 3rd African Human-Computer Interaction Conference: Inclusiveness and Empowerment},
  pages={52--64},
  year={2021}
}

@inproceedings{velykoivanenko2024designing,
  title={Designing a Data-Driven Survey System: Leveraging Participants' Online Data to Personalize Surveys},
  author={Velykoivanenko, Lev and Salehzadeh Niksirat, Kavous and Teofanovic, Stefan and Chapuis, Bertil and Mazurek, Michelle L and Huguenin, K{\'e}vin},
  booktitle={Proceedings of the CHI Conference on Human Factors in Computing Systems},
  pages={1--22},
  year={2024}
}

@inproceedings{ebert2023qbutterfly,
  title={QButterfly: Lightweight Survey Extension for Online User Interaction Studies for Non-Tech-Savvy Researchers},
  author={Ebert, Nico and Scheppler, Bj{\"o}rn and Ackermann, Kurt Alexander and Geppert, Tim},
  booktitle={Proceedings of the 2023 CHI Conference on Human Factors in Computing Systems},
  pages={1--8},
  year={2023}
}

@article{molnar2019smartriqs,
  title={SMARTRIQS: A simple method allowing real-time respondent interaction in Qualtrics surveys},
  author={Molnar, Andras},
  journal={Journal of Behavioral and Experimental Finance},
  volume={22},
  pages={161--169},
  year={2019},
  publisher={Elsevier}
}

@article{zarouali2024comparing,
  title={Comparing chatbots and online surveys for (longitudinal) data collection: an investigation of response characteristics, data quality, and user evaluation},
  author={Zarouali, Brahim and Araujo, Theo and Ohme, Jakob and de Vreese, Claes},
  journal={Communication Methods and Measures},
  volume={18},
  number={1},
  pages={72--91},
  year={2024},
  publisher={Taylor \& Francis}
}

@article{rhim2022application,
  title={Application of humanization to survey chatbots: Change in chatbot perception, interaction experience, and survey data quality},
  author={Rhim, Jungwook and Kwak, Minji and Gong, Yeaeun and Gweon, Gahgene},
  journal={Computers in Human Behavior},
  volume={126},
  pages={107034},
  year={2022},
  publisher={Elsevier}
}

@inproceedings{wen2022hybrid,
  title={Hybrid Online Survey System with Real-Time Moderator Chat},
  author={Wen, James and Colley, Ashley},
  booktitle={Proceedings of the 21st International Conference on Mobile and Ubiquitous Multimedia},
  pages={257--258},
  year={2022}
}

@article{xiao2020tell,
  title={Tell me about yourself: Using an AI-powered chatbot to conduct conversational surveys with open-ended questions},
  author={Xiao, Ziang and Zhou, Michelle X and Liao, Q Vera and Mark, Gloria and Chi, Changyan and Chen, Wenxi and Yang, Huahai},
  journal={ACM Transactions on Computer-Human Interaction (TOCHI)},
  volume={27},
  number={3},
  pages={1--37},
  year={2020},
  publisher={ACM New York, NY, USA}
}

@article{reips2008interval,
  title={Interval-level measurement with visual analogue scales in Internet-based research: VAS Generator},
  author={Reips, Ulf-Dietrich and Funke, Frederik},
  journal={Behavior research methods},
  volume={40},
  number={3},
  pages={699--704},
  year={2008},
  publisher={Springer}
}

@article{joshi2015likert,
  title={Likert scale: Explored and explained},
  author={Joshi, Ankur and Kale, Saket and Chandel, Satish and Pal, D Kumar},
  journal={British journal of applied science \& technology},
  volume={7},
  number={4},
  pages={396--403},
  year={2015}
}

@article{gentzkow2019text,
  title={Text as data},
  author={Gentzkow, Matthew and Kelly, Bryan and Taddy, Matt},
  journal={Journal of Economic Literature},
  volume={57},
  number={3},
  pages={535--574},
  year={2019},
  publisher={American Economic Association 2014 Broadway, Suite 305, Nashville, TN 37203-2425}
}

@article{wolfers2004prediction,
  title={Prediction markets},
  author={Wolfers, Justin and Zitzewitz, Eric},
  journal={Journal of economic perspectives},
  volume={18},
  number={2},
  pages={107--126},
  year={2004},
  publisher={American Economic Association}
}

@inproceedings{kriplean2012supporting,
  title={Supporting reflective public thought with considerit},
  author={Kriplean, Travis and Morgan, Jonathan and Freelon, Deen and Borning, Alan and Bennett, Lance},
  booktitle={Proceedings of the ACM 2012 conference on Computer Supported Cooperative Work},
  pages={265--274},
  year={2012}
}

@book{behrens2014principles,
  title={The principles of LiquidFeedback},
  author={Behrens, Jan and Kistner, Axel and Nitsche, Andreas and Swierczek, Bj{\"o}rn},
  year={2014},
  publisher={Interacktive Demokratie}
}

@article{salganik2015wiki,
  title={Wiki surveys: Open and quantifiable social data collection},
  author={Salganik, Matthew J and Levy, Karen EC},
  journal={PloS one},
  volume={10},
  number={5},
  pages={e0123483},
  year={2015},
  publisher={Public Library of Science San Francisco, CA USA}
}

@article{boone2005rigour,
  title={Rigour in quantitative analysis: The promise of Rasch analysis techniques},
  author={Boone, William and Rogan, John},
  journal={African Journal of Research in Mathematics, Science and Technology Education},
  volume={9},
  number={1},
  pages={25--38},
  year={2005},
  publisher={Taylor \& Francis}
}

@misc{polis,
  title = {{Pol.is}},
  author = {{Pol.is}},
  year = {2015},
  url = {https://pol.is},
  note = {Collaborative opinion-gathering platform},
}

@article{cypress2019data,
  title={Data analysis software in qualitative research: Preconceptions, expectations, and adoption},
  author={Cypress, Brigitte S},
  journal={Dimensions of critical care nursing},
  volume={38},
  number={4},
  pages={213--220},
  year={2019},
  publisher={LWW}
}

@article{castleberry2018thematic,
  title={Thematic analysis of qualitative research data: Is it as easy as it sounds?},
  author={Castleberry, Ashley and Nolen, Amanda},
  journal={Currents in pharmacy teaching and learning},
  volume={10},
  number={6},
  pages={807--815},
  year={2018},
  publisher={Elsevier}
}

@article{dai2023llm,
  title={LLM-in-the-loop: Leveraging large language model for thematic analysis},
  author={Dai, Shih-Chieh and Xiong, Aiping and Ku, Lun-Wei},
  journal={arXiv preprint arXiv:2310.15100},
  year={2023}
}

@article{chamberlin1983representative,
  title={Representative deliberations and representative decisions: Proportional representation and the Borda rule},
  author={Chamberlin, John R and Courant, Paul N},
  journal={American Political Science Review},
  volume={77},
  number={3},
  pages={718--733},
  year={1983},
  publisher={Cambridge University Press}
}

@inproceedings{baumer2017subjects,
  title={When subjects interpret the data: Social media non-use as a case for adapting the delphi method to cscw},
  author={Baumer, Eric PS and Xu, Xiaotong and Chu, Christine and Guha, Shion and Gay, Geri K},
  booktitle={Proceedings of the 2017 acm conference on computer supported cooperative work and social computing},
  pages={1527--1543},
  year={2017}
}

@inproceedings{yu2015can,
  title={Can survey instructions relieve respondent burden},
  author={Yu, Erica C and Fricker, Scott and Kopp, Brandon},
  booktitle={70th Annual Conference of the American Association for Public Opinion Research, Hollywood, FL},
  year={2015}
}

@book{beatty2019advances,
  title={Advances in questionnaire design, development, evaluation and testing},
  author={Beatty, Paul C and Collins, Debbie and Kaye, Lyn and Padilla, Jose-Luis and Willis, Gordon B and Wilmot, Amanda},
  year={2019},
  publisher={John Wiley \& Sons}
}

@article{charmaz2015grounded,
  title={Grounded theory},
  author={Charmaz, Kathy},
  journal={Qualitative psychology: A practical guide to research methods},
  volume={3},
  pages={53--84},
  year={2015}
}

@article{braun2019reflecting,
  title={Reflecting on reflexive thematic analysis},
  author={Braun, Virginia and Clarke, Victoria},
  journal={Qualitative research in sport, exercise and health},
  volume={11},
  number={4},
  pages={589--597},
  year={2019},
  publisher={Taylor \& Francis}
}

@article{baum2006participatory,
  title={Participatory action research},
  author={Baum, Fran and MacDougall, Colin and Smith, Danielle},
  journal={Journal of epidemiology and community health},
  volume={60},
  number={10},
  pages={854},
  year={2006}
}

@inproceedings{kinsella2006hermeneutics,
  title={Hermeneutics and critical hermeneutics: Exploring possibilities within the art of interpretation},
  author={Kinsella, Elizabeth Anne and others},
  booktitle={Forum Qualitative Sozialforschung/Forum: Qualitative Social Research},
  volume={7},
  number={3},
  year={2006}
}

@article{elwood1998gis,
  title={GIS and community-based planning: Exploring the diversity of neighborhood perspectives and needs},
  author={Elwood, Sarah and Leitner, Helga},
  journal={Cartography and Geographic Information Systems},
  volume={25},
  number={2},
  pages={77--88},
  year={1998},
  publisher={Taylor \& Francis}
}

@article{schugurensky2024participatory,
  title={Participatory budgeting and local development: Impacts, challenges, and prospects},
  author={Schugurensky, Daniel and Mook, Laurie},
  journal={Local Development \& Society},
  volume={5},
  number={3},
  pages={433--445},
  year={2024},
  publisher={Taylor \& Francis}
}

@article{okada2013community,
  title={Community-based decision making in Japan},
  author={Okada, Norio and Fang, Liping and Kilgour, D Marc},
  journal={Group Decision and Negotiation},
  volume={22},
  pages={45--52},
  year={2013},
  publisher={Springer}
}

@article{mcdavitt2016dissemination,
  title={Dissemination as dialogue: building trust and sharing research findings through community engagement},
  author={McDavitt, Bryce and Bogart, Laura M and Mutchler, Matt G and Wagner, Glenn J and Green Jr, Harold D and Lawrence, Sean Jamar and Mutepfa, Kieta D and Nogg, Kelsey A},
  journal={Preventing Chronic Disease},
  volume={13},
  pages={E38},
  year={2016}
}

@article{montoya2011dialogical,
  title={Dialogical action: Moving from community-based to community-driven participatory research},
  author={Montoya, Michael J and Kent, Erin E},
  journal={Qualitative Health Research},
  volume={21},
  number={7},
  pages={1000--1011},
  year={2011},
  publisher={SAGE Publications Sage CA: Los Angeles, CA}
}

\appendix

\section{Prompts}
\subsection{Clustering prompt}\label{clustering prompt}
\begin{verbatim}
I’m running a survey that involves qualitative data analysis. I’d like your help 
extracting insightful theme(s) from each response as they come in. Can you identify 
all the new/existing theme(s) that the response falls into. I will provide you with 
the current list of extracted themes that were derived from the previous responses 
as well as the new response.

The survey question I asked is: {survey question}

A couple of guidelines to keep in mind:
1. Themes are general categories that represent shared ideas between responses. You 
must strike a balance and create a theme that is clear, concise, and useful for 
 thematic analysis.
2. Theme names should be at least 3 words long. Someone should be able to understand 
the core aspect of the theme by only reading the theme name.
3. Any themes generated should be fully unique with no overlap between them. 
Redundant themes are not useful and responses should be placed in existing themes 
before new themes are generated.
4. Responses can be in 1 or many themes.
5. Try to minimize the number of themes generated while still capturing the IMPORTA-
NT aspects of the response.

I’d like you to think step-by-step using the following strategy:
1. Search the existing themes and find any themes the current response CLEARLY belo-
ngs to.
2. If you believe that the current themes do not properly capture ALL core ideas of 
the response, create a new general theme or multiple new themes that fully capture 
the ideas in the response.
3. Double check all themes are unique and that the response is fully represented by 
the assigned themes. If any of the new themes are overlapping with an existing theme, 
place the response in the existing theme first.
4. Return the list of themes for the response in a comma separated list

Please ONLY report your final decision in a comma separated list of themes.

Here are the current themes:
{list of cluster titles and descriptions} 

Here is the new response: {response}
\end{verbatim}

\subsection{Follow-up prompt}
\label{follow-up prompt}\begin{verbatim} 
We asked the following survey question: {survey question}
Here is the response we received from the user: {new response}
Please analyze the response and create a customized follow-up question for the 
respondent that encourages them to think deeper about their response and provide the 
survey creator with deeper insights. Only output the question.

\end{verbatim}

\end{document}